\RequirePackage{fix-cm}
\documentclass[smallextended]{svjour3}

\smartqed  % flush right qed marks, e.g. at end of proof

\usepackage{graphicx}
\usepackage{tabularx}
\usepackage{amsmath}
\usepackage{comment}
\usepackage[colorlinks,allcolors=green!60!black]{hyperref}
\usepackage{booktabs}
\usepackage{mathtools}
\usepackage{dirtytalk}

\newcommand{\Ft}{\mathbf{F}}

\newcommand{\Tt}{\mathbf{T}}
\newcommand{\It}{\mathbf{I}}

\newcommand{\Dt}{\mathbf{d}}
\newcommand{\Lt}{\mathbf{l}}
\renewcommand{\Tt}{\mathbf{t}}

\newcommand{\ve}[1]{\boldsymbol{#1}}
\newcommand{\te}[1]{\boldsymbol{#1}}

\renewcommand{\div}{\operatorname{div}\,}

\newcommand{\sigmat}{\boldsymbol{\sigma}}
\newcommand{\omegat}{\boldsymbol{\omega}}

\newcommand{\Jt}{\mathbf{J}}
\newcommand{\vt}{\mathbf{v}}
\newcommand{\Vt}{\mathbf{V}}
\newcommand{\nt}{\vec{n}}

\newcommand{\xt}{\mathbf{x}}
\newtheorem{algo}{Algorithm}

\usepackage{amsmath,amssymb,dsfont,mathtools}
\usepackage{graphicx}
\usepackage{xcolor}
\usepackage{units}
\usepackage{pgfplots}
\usepackage{siunitx}
\usepackage{tabularx}

\begin{document}

\setcounter{tocdepth}{3}

\title{A finite element / neural network framework for modeling
  suspensions of non-spherical particles}
\subtitle{Concepts and medical applications}

\titlerunning{Non-spherical particles in medical flow problems}  

\author{Martyna Minakowska \and Thomas Richter  \and Sebastian Sager}

\institute{M. Minakowska and T. Richter \at
	Institute of Analysis and Numerics, Otto-von-Guericke
	University Magdeburg, \\ Universit\"atsplatz 2, 
	39106 Magdeburg, Germany,\\
	\email{\{martyna.minakowska, thomas.richter\}@ovgu.de}           %  \\
	\and
	S. Sager \at
	Institute of Mathematical Optimization, Otto-von-Guericke
	University Magdeburg, \\ Universit\"atsplatz 2, 39106 Magdeburg,
	Germany,\\
	\email{sebastian.sager@ovgu.de}
}

\date{Received: date / Accepted: date}

\maketitle

\begin{abstract}
	
An accurate prediction of the translational and rotational motion of particles suspended 
in a fluid is only possible if a complete set of correlations for the force coefficients 
of fluid-particle interaction is known. The present study is thus devoted to the derivation 
and validation of a new framework to determine the drag, lift, rotational and pitching 
torque coefficients for different non-spherical particles in a fluid flow. The motivation 
for the study arises from medical applications, where particles may have an arbitrary 
and complex shape. Here, it is usually not possible to derive accurate analytical models 
for predicting the different hydrodynamic forces. However, considering for example the 
various components of blood, their shape takes an important role in controlling several 
body functions such as control of blood viscosity or coagulation. Therefore, the presented 
model is designed to be applicable to a broad range of shapes. Another important feature 
of the suspensions occurring in medical and biological applications is the high number 
of particles. The modelling approach we propose can be efficiently used for simulations 
of solid-liquid suspensions with numerous particles. Based on resolved numerical simulations 
of prototypical particles we generate data to train a neural network which allows us 
to quickly estimate the hydrodynamic forces experienced by a specific particle immersed in a fluid.


\end{abstract}

\section{Introduction}\label{intro}

The prediction of the motion of non-spherical particles suspended 
in a fluid is crucial for the understanding of natural processes 
and industrial applications. In such processes, particles can have 
different shapes and sizes, may be deformed and can interact with 
each other. So far, in the majority of scientific studies, 
particulate flow modelling is investigated with the hypothesis 
of perfectly spherical particles, thereby eliminating orientation 
and shape effects. This assumption is very convenient due to its 
simplicity, the fact that the behaviour of spheres is well known 
and the availability of a number of models to describe the interaction 
with fluid flow. The study of suspensions of multiple, 
irregular-shaped, interacting and deformable particles has 
received less attention and still presents a challenge.

Particles come in all sort of shapes and sizes, in fact, 
due to the arbitrary nature of naturally occurring particles 
there are an indefinite number of possible shapes. On the 
other hand, there is a common understanding that particle shape 
has a strong influence on the dynamics of NSPS (non-spherical 
particulate systems). These two factors combined makes modelling 
of NSPS in general way impossible, since for describing the motion 
of non-spherical particles, detailed information on the fluid dynamic 
forces acting on such particles are necessary, but generally 
not available. Therefore, particular 
models place emphasis on different shapes and types of flow. 
In our work we focus on medical applications, namely on modelling 
of platelets dynamics under blood flow. Platelets play a main role 
in the process of blood coagulation and therefore are of great interest 
in the modelling of blood flow. The majority of  models characterize 
platelet motion quantitatively and use approaches such as 
immersed boundary method~\cite{Fogelson2008}, cellular Potts 
model~\cite{Xu2007,Xu2008,Xu2010} and dissipative particle 
dynamics~\cite{Filipovic2008,Tosenberger2013}, treating platelets 
as points and thus neglecting entirely their shape. Some effort 
has been done to model platelets as rigid two- or three-dimensional 
spheres or spheroids~\cite{King2005,Sweet2011,Slepian2014} but this 
simplification of shape has been shown to affect processes in which 
platelets are involved (e.g. spherical platelets marginate faster 
than ellipsoidal and disc forms)~\cite{2019_Book_Microflows}. 

On the other hand, several studies have been performed to model 
particles of irregular shapes, but motivation behind them usually 
arises from engineering applications (dispersion of pollutant, 
pulverized coal combustion, pneumatic transport)~\cite{Zhong2016}, 
where particles are much bigger and usually constitute a significant 
part of the volume of the suspension~\cite{Shardt2012}. But even 
under given very specific circumstances there is no set of correlations 
of the forces acting on irregular-shaped particles suspended in a fluid 
(forces arising from fluid-particle interaction). Furthermore, 
in these kinds of models interactions between particles become dominant 
in terms of determining particle dynamics, whereas platelets are very 
dilute in blood and their contact is rather rare. 

The motivation behind this work is fourfold and arises from the 
specificity of the modelled phenomenon. Firstly, platelets, 
because of their small size (compared to blood vessel), are often 
modelled quantitatively, ignoring the importance of shape effects. 
Secondly, platelets constitute only a very small volume of blood 
what makes most of the engineering applications-driven models, 
where particles are very dense and often interact with each other, 
inappropriate. Thirdly, highly irregular shape of platelets requires 
new methods for estimating force coefficients in fluid-particle interaction.
Fourthly, platelets are numerous and the evolution of their movement
needs to be evaluated quickly and efficiently.

A common approach to the problem of modelling of NSPS consists in 
developing analytical models for fluid dynamic forces acting 
on particles, cf.~\cite{Holzer2008,Lu2015,Rosendahl2010,Wachem2012,Zhong2016}. 
In this contribution we go a different way and refrain from giving
analytical expressions. Instead, by prototypical simulations 
we train a neural network model that
takes a couple of parameters describing the particle shape, size and the
flow configuration as input and which gives hydrodynamical
coefficients like drag, lift and torque as output. We demonstrate that
such a model can be efficiently trained in an offline phase for a
range of particles. Lateron, the coupling of the flow model with the
particle system only requires the evaluation of the network for getting
updates on these coefficients. This two-step approach with an offline phase for training a network based on the particle classes under consideration allows for a direct extension to further applications.

A similar approach is considered by in~\cite{YanHeTangWang2019}. Here,
the authors design and train a radial basis function network to
predict the drag coefficient of non-spherical particles in fluidized
beds. Training is based on experimental data and the network input is
the particle's sphericity and the Reynolds number, covering the Stokes
and the intermediate regime. Our approach, based on training data
generated by detailed simulations of prototypical particles for
predicting drag, lift and torque coefficients could by augmented by
including further experimental data.

In the following section we will
describe prototypical medical applications where such a heterogenous
modeling approach can be applied. Then, in Section~\ref{sec:model} we
detail the general framework for coupling the Navier-Stokes equation
with a discrete particle model. Section~\ref{sec:dnn} introduces
the neural network approach for estimating the hydrodynamical
coefficients and we describe the procedure for offline training of the
network. Then, different numerical test cases are described in
Section~\ref{sec:results} that show the potential of such a heterogeneous
modeling approach. Section~\ref{sec:sim} is devoted to a numerical
study comprising many particles and shows the efficiency of the presented approach.
We summarize with a short conclusion in Section~\ref{sec:conc}.

\section{Modeling of suspensions with non-spherical particles and
  medical applications}\label{sec:overview}

Platelets are a vital component of the blood clotting mechanism. 
They are small non-nucleated cell fragments. They have a diameter 
of approximately $\unit[2-4]{\mu m}$, thickness of $\unit[0.5]{\mu m}$, 
volume of about $\unit[7]{\mu m^3}$ and a number density 
of $\unit[1.5-4\cdot10^5]{\mu l^{-1}}$ \cite{2017_Book_Hemomath} which leads to a volume fraction of only about $10^{-3}:1$.  
Still they are a vital component of the blood clotting mechanism. 
In the rest state platelets shape is discoid, but they have the ability 
of deforming as a response to various stimuli (chemical and mechanical). 
They may become star shaped (rolling over blood vessels wall to inspect 
its integrity). During the clotting process they undergo deep 
morphological changes, from becoming spherical and emitting 
protuberances (philopodia or pseudopods) which favour mutual 
aggregation, as well as adhesion to other elements constituting 
the clot to fully flat, spread stage, to enable wound closure. 
Thrombocytes constitute approximately less than $1\%$ of the 
blood volume, therefore individual platelets have a negligible 
effect on blood rheology \cite{Fogelson2015}. 

Due to a significant effect of the particles shape on their motion 
and their practical importance in industrial applications, the 
non-sphericity started attracting attention in the modelling 
and simulations of particles transport in fluid 
flows~\cite{Holzer2009,Rosendahl2010,Wachem2012}. Unfortunately,
it is not possible to consider each shape in the implementation
of numerical methods because of non-existence of a single approach 
describing accurately the sizes and shapes of non-spherical particles.
Spheres can be described by a single characteristic value, 
i.e. the diameter, whereas non-spherical particles require more parameters.
Even very regular shapes need at least two parameters. Moreover, 
the particles may have varying orientation with respect to the flow, 
what makes the description of their behaviour even more difficult. 
Even though several methods for shape parametrization and measurement 
have been suggested, none has won greater acceptance. 
One of most commonly used shape factor is a sphericity  
which was firstly introduced in~\cite{Wadell1934} and defined as 
the ratio between the surface area of a sphere with equal volume 
as the particle and the surface area of the considered particle. 
Then the drag coefficient for non-spherical particle is estimated by using 
correlations for spherical particles and modified to take into 
account the sphericity factor~\cite{Yow2005}. Models using sphericity 
as a shape factor give promising results when restricted to non-spherical particles 
with aspect ratios $\beta$ less than 1.7~\cite{1978_Book_BubblesDropsAndParticles}, 
where $\beta = L/D$ with $L$ and $D$ being length and diameter 
of the considered particle. For particles having extreme shapes 
and those having little resemblance to a sphere, the sphericity 
concept fails to produce satisfying quantitative results~\cite{Chhabra1999}. 
In general, the lower the sphericity, the poorer is the prediction.
Also, the same value of the sphericity may be obtained for two very 
varying shapes whose behaviour in the flow is different. 
Moreover, the sphericity does not take the orientation into account. 
In order to introduce orientation dependency in drag correlations, 
some researchers use two additional factors: the crosswise sphericity 
and lengthwise sphericity~\cite{Holzer2008}. 
Most of these correlations employ also dependency on particle Reynolds 
number defined as
\[
Re_p = \rho\bar u d_{eq}/\mu,
\] 
where $\rho$ and $\mu$ are fluid density and viscosity, $\bar u =
u_f-u_p$ is the velocity  of the particle relative to the fluid
velocity and $d_{eq}$ is the  equivalent particle diameter, i.e. the
diameter of a sphere  with the same volume as the considered particle
in order to  include the importance of fluid properties. 

The shape factor concept may be described as an attempt 
to define a single correlation for drag for all shapes and orientations. 
Another approach appeared as an alternative 
consisting in obtaining drag coefficient expressions for a fixed 
shape and any orientations: the drag coefficient is determined 
at two extreme angles of incidence ($\ang{0}$ and $\ang{90}$) 
from existing correlations which are then linked by some functions
resulting in the whole range of angles of incidence for non-spherical 
particles~\cite{Rosendahl2000}. However, besides drag force, 
non-spherical particles are associated with orientation 
and shape induced lift along with pitching and rotational torques.
H\"{o}lzer and Sommerfeld~\cite{Holzer2009} investigated a few
different shapes of non-spherical particles at different flow incident angles
using the Lattice Boltzmann method to simulate the flow around the particle.
Wachem et al.~\cite{Wachem2012} proposed new force correlations 
(for drag, lift, pitching and rotational torque) for particular shapes 
of non-spherical particles (two ellipsoids with different aspect ratio, disc and fibre) 
from data given by a direct numerical simulation (DNS) carried out 
with an immersed boundary method. Those correlations employ
particle Reynolds number, angle of incidence and some shape-related coefficients. 
Ouchene et al.~\cite{Ouchene2016} determined 
force coefficients depending on particle Reynolds number, 
aspect ratio and angle of incidence by fitting the results extracted 
from DNS computations of the flow around prolate ellipsoidal particles.
Discrete element methods (DEM) coupled with computational 
fluid dynamics (CFD) has been recognized as a promising method 
to meet the challenges of modelling of NSPS~\cite{Zhu2007,Zhu2008}. 
DEM is a numerical approach for modelling a large number of particles 
interacting with each other.  
The simplest computational sequence for the DEM typically 
proceeds by solving the equations of motion, while updating 
contact force histories as a consequence of contacts 
between different discrete elements and/or resulting 
from contacts with model boundaries. It is designed to deal
with very dense suspensions, where contacts between particles
are very common and play a key role in determining the motion of particles, 
see~\cite{Zhong2016} for an extensive overview of DEM. 

Obviously, for a particle with a certain specific shape,
the general expressions derived from the first factor shape 
approach tend to be less accurate than the specialized one 
for that shape, but the efficiency of interpolations/extrapolations 
to the various shapes to provide the general expression is an attractive 
perspective on engineering applications.
On the other hand, particles occurring in biological processes 
are usually very numerous. Therefore, an effective method not only 
has to be accurate but also efficient in terms of computational time.

To overcome the aforementioned limitations in modelling of NSPS
we employ the recently trending approach  and use machine learning 
to design a method that enables us to model the behaviour of
suspensions of particles of an arbitrary shape while 
maintaining at the same time the accuracy of shape-specific models. 
We also place an emphasis on the computational efficiency as usually 
there are plenty of particles involved in medical processes 
and engineering problems.

\section{Model description}\label{sec:model}

This section describes the general numerical framework for suspensions of particles in a Navier-Stokes fluid. The discretization of the Navier-Stokes equations is realized in the finite element toolbox \emph{Gascoigne 3D}~\index{gascoigne3d} and outlined in Section~\ref{sec:fluid}. Then, in Section~\ref{sec:particles} we describe a very simple model for the motion of the particles.

\subsection{Fluid dynamics}\label{sec:fluid}

Consider  a  finite  time interval $I= [0,T]$  and  a  bounded  domain  $\Omega \in \mathbb{R}^d$ for $d\in\{2,3\}$. 
We assume incompressibility of fluid, which is modelled by the Navier-Stokes equations that take the form
\begin{equation}\label{ns}
  \begin{aligned}
    \rho\big(\partial_t\vt + (\vt \cdot\nabla)\vt\big) - \div
    \sigmat(\vt,p)= 0, \\     \div \vt = 0,
  \end{aligned}
\end{equation}
where $\vt$ denotes the fluid velocity, $\sigmat$ is the Cauchy stress tensor
\[
\sigmat(\vt,p)=\rho\nu(\nabla\vt+\nabla\vt^T)-p\It,
\]
$p$ the pressure, $\rho$ the fluid mass density and  $\nu$ the kinematic viscosity. Fluid density and viscosity are assumed to be nonnegative and constant.

The  fluid  boundary  is  split  into an inflow boundary $\Gamma_{in}$,  an outflow boundary $\Gamma_{out}$ and rigid no-slip wall boundaries $\Gamma_{wall}$. 
On the inflow and walls we impose Dirichlet boundary conditions while on the outflow we apply the do-nothing condition (see e.g. \cite{Heywood1996})
\begin{equation}\label{bc}
  \begin{aligned}
    \vt	&=\vt_{in}& \text{ on }  &\Gamma_{in}\times I, \\
    \vt	&=0 	 &  \text{ on } & \Gamma_{wall}\times I, \\
    (\rho\nu\nabla\vt-p\It)\nt &= 0 &	   \text{ on } & \Gamma_{out}\times I,
  \end{aligned}
\end{equation}
where $\vt_{in}$ is prescribed inflow-profile and $\nt$ is the outward unit normal vector.

\paragraph{Discretization}

For temporal discretization of the Navier-Stokes equations we introduce a uniform partitioning of the interval $I=[0,T]$ into discrete steps
\[
0=t_0<t_1<\dots<t_N=T,\quad k\coloneqq t_n-t_{n-1}. 
\]
By $\vt_n\coloneqq \vt(t_n)$ and $p_n\coloneqq p(t_n)$ we denote the approximations at time $t_n$. We use a shifted version of the Crank-Nicolson time discretization scheme which is second order accurate and which has preferable smoothing properties as compared to the standard version, see~\cite[Remark 1]{HeywoodRannacher1990}, i.e.
\begin{multline}\label{shiftedCN}
  \rho \left(\frac{\vt_n-\vt_{n-1}}{k} +
  \theta (\vt^n\cdot\nabla)\vt^n
  +(1-\theta) (\vt^{n-1}\cdot\nabla)\vt^{n-1}
  \right)\\
  -\theta \div \sigmat(\vt^n,p^n)
  -(1-\theta) \div \sigmat(\vt^n,p^n)
  = 0,
\end{multline}
where, typically, $\theta=\frac{1+k}{2}$. For spatial discretization we denote by $\Omega_h$ a quadrilateral (or hexahedral) finite element mesh of the domain $\Omega$ that satisfies the usual regularity assumptions required for robust interpolation estimates, see~\cite[Section 4.2]{Richter2017}. Adaptive meshes are realized by introducing at most one hanging node per edge. Discretization is based on second order finite elements for pressure and velocity. To cope with the lacking inf-sup stability of this equal order finite element pair we stabilize with the local projection method~\cite{BeckerBraack2001}. Local projection terms are also added to stabilize dominating transport~\cite{BeckerBraack2004}. Finally, velocity $\vt_n\in [V_h]^2$ and pressure $p_n\in V_h$ (where we denote by $V_h$ the space of bi-quadratic finite elements on the quadrilateral mesh) are given as solution to 
\begin{multline}\label{discreteNS}
  \big(\rho \vt_n,\phi_h\big)_\Omega
  +k \theta \big( \rho (\vt_n\cdot\nabla)\vt_n,\phi_h\big)_\Omega
  +k \theta \big(\rho\nu(\nabla\vt_n+\nabla\vt_n^T),\nabla\phi\big)_\Omega\\ 
  -k \big(p_n,\div\,\phi\big)_\Omega
  + k\big(\rho\div\vt_n,\xi_h\big)_\Omega \\
  +  k\sum_{K\in\Omega_h} \alpha_K \big(
  \nabla (p_n-\pi_hp_n),\nabla (\xi_h-\pi_h \xi_h)\big)_K\\
   +k \sum_{K\in\Omega_h} \alpha_K \big(
  (\vt_n\cdot\nabla)(\vt_n-\pi_h\vt_n),(\vt_n\cdot\nabla) (\phi_h-\phi^\xi_h)\big)_K\\ 
  =
  \big(\rho\vt_{n-1},\phi_h\big)_\Omega
  -k (1-\theta) \big( \rho (\vt_{n-1}\cdot\nabla)\vt_{n-1},\phi_h\big)_\Omega\\
  -k (1-\theta) \big(\rho\nu(\nabla\vt_{n-1}+\nabla\vt_{n-1}^T),\nabla\phi\big)_\Omega\quad\forall (\phi_h,\xi_h)\in [V_h]^2\times V_h,
\end{multline}
where the stabilization parameters are element-wise chosen as~\cite{BraackBurmanJohnEtAl2007}
\[
\alpha_K =\alpha_0 \left(\frac{\nu}{h_K^2}
+\frac{\|\vt\|_{L^\infty(K)}}{h_K} + \frac{1}{k}\right)^{-1},
\]
where $h_K=\operatorname{diam}(K)$ is the diameter of the element $K$. Usually we choose  $\alpha_0=0.1$. By $\pi_h:V_h\to V_h^{(1)}$ we denote the interpolation into the space of bi-linear elements on the same mesh $\Omega_h$.

\paragraph{Solution of the discretized problem}

Discretization by means of~(\ref{discreteNS}) gives rise to a large system of nonlinear algebraic equation which we approximate by a Newton scheme based on the analytical Jacobian of~(\ref{discreteNS}). The resulting linear systems are solved by a GMRES iteration (Generalized minimal residual method~\cite{Saad1996}), preconditioned by a geometric multigrid solver~\cite{BeckerBraack2000a}. As smoother we employ a Vanka-type iteration based on the inversion of the submatrices belonging to each finite element cell. These local $27\times 27$ ($108\times 108$ in 3d) matrices are inverted exactly. Essential parts of the complete solution framework are parallelized using OpenMP, see~\cite{FailerRichter2020}.

\subsection{Particle dynamics}\label{sec:particles}

The particles suspended in the fluid are described as rigid bodies and their dynamics is driven by the hydrodynamical forces of the flow. Each particle $P$ with center of mass $\xt_P$, velocity $\vec V_P$ and angular velocity $\vec \Omega_P$ is governed by Newton's law of motion 
\begin{equation}\label{motion2d}
  \begin{aligned}
    m_P \vec X_p''(t) &= \vec F(\vt,p;P)\coloneqq -\int_{\partial P}
    \sigmat(\vt,p)\nt_P\,\text{d}s\\
    \Jt_P \vec\Omega_P'(t) &= \vec T(\vt,p;P)\coloneqq \int_{\partial P}
    (\xt-\xt_P)\times \big(\sigmat(\vt,p)\nt_P\big)\,\text{d}s, 
  \end{aligned}
\end{equation}
where $m_P$ is the particle's mass, and $\Jt_P$ its moment of inertia
given by
\[
\Jt_P = \operatorname{diag}\Big\{
\rho_P\int_P\Big(
|\xt-\xt_P|^2 - (\xt_i-[\xt_P]_i)^2
\Big) \,\text{d}\xt; i=1,2,3\Big\},
\]
with the (uniform) particle density $\rho_P$. 
$\nt_P$ is the unit normal vector on the
particle boundary facing into the fluid.

A resolved simulation is out of bounds due to the large number of
platelets and in particular due to the discrepancy in particle diameter (about $\unit[10^{-6}]{m}$) versus vessel
diameter (about $\unit[10^{-3}]{m}$). Instead, we consider all platelets to be point-shaped and determine traction forces $\vec F(\vt,p;P)$ and torque $\vec
T(\vt,p;P)$ based on previously trained neural networks. These coefficients will
depend on the shape and the size of the particles but also on their
relative orientation and motion in the velocity field of the blood. 
Since the relative
velocities (blood vs. particles) are very small the interaction lies
within the Stokes regime with a linear scaling in terms of the
velocity. The deep neural network will predict coefficients for
determining coefficients of drag $C_d$, lift $C_l$, pitching torque
$C_p$ and rotational torque $C_r$ and the resulting forces exerted on
each particle $P$ are given by
\begin{equation}\label{forces:coefficients}
  \begin{aligned}
    \vec F_P &=
    C_d(P,\psi_P)(\vt-\vec V_p)
    +C_l(P,\psi_P)(\vt-\vec V_p)^\perp\\
    \vec T_P &=
    C_p(P,\psi_P)|\vt-\vec V_p|
    +C_r(P,\psi_P)(\omegat-\vec \Omega_p)
  \end{aligned},
\end{equation}
where $P=(L_x,L_y,L_z,\alpha_{top},\alpha_{bot})$ describes the
particle shape 
and where $\psi_P$ is the relative angle of attack which depends on
the particle orientation but also on the relative velocity vector between
blood velocity and particle trajectory, see
Figure~\ref{fig:angle}. The coefficent functions $C_d,C_l,C_p$ and
$C_r$ will be trained based on detailed numerical simulations using 
random  particles in random configurations. By $(\vt-\vec V_p)^\perp$
we denote the flow vector in lift-direction, orthogonal to the main
flow direction. In 3d configurations, two such lift coefficients must
be trained. Here we will however only consider 2d simplifications with
one drag and one lift direction.

\newpage
\section{An artificial neural network model for predicting hydrodynamical
  parameters}\label{sec:dnn}

In this section we describe the neural network model for coupling the Navier-Stokes equation with a suspension of non-spherical particles. The different hydrodynamical coefficients will be taken from a neural network, which is trained in an offline phase. Training data is achieved by resolved Navier-Stokes simulations using prototypical particles with random parameters. 

The setting investigated in this work carries several special characteristics that differ from industrial applications. 
\begin{itemize}
\item The particle density is very small - about $\unit[1.04-1.08]{\cdot 10^3 gl^{-1}}$
  and the particle and fluid densities are similar (the average density of 
  whole blood for a human  is about $\unit[1.06 \cdot 10^3]{gl^{-1}}$).
  Blood contains about 200\,000 -- 400\,000 platelets per $\unit{mm}^3$ summing up
  to less than 1\%  of the overall blood volume \cite{Wiwanitkit2004}. 
  Hence we neglect
  all effects of the particles onto the fluid. This
  simplification is possible since we only model the platelets as
  rigid particles. Effects of the red blood cells,
  much larger and appearing in greater quantity, can be integrated by means of a
  non-Newtonian rheology.
\item The particle Reynolds numbers are very small (with order of
  magnitude about  $10^{-4}$ or less) such that we are locally in the
  Stokes regime. This is  mainly due to the smallness of the platelets
  (diameter approximately  $\unit[3]{\mu m}$) and the small flow
  velocities at (bulk) Reynolds  numbers ranging from $50$ to $1\,000$
  depending on the specific
  vessel under investigation. We focus on coronary vessels with a diameter 
  around $\unit[2]{mm}$ and with Reynolds number about 200. 
\item The platelets have a strongly non-spherical, disc-like
  shape. Their shape and size underlies a natural variation. Furthermore, under activation, the particles will take a spherical shape. 
\end{itemize}
Instead of deriving analytical models for the transmission of forces
from the fluid to the particles, we develop a neural network for the
identification of drag, lift and torque coefficients based on several
parameters describing the shape and the size of the platelets and the
individual flow configuration.

\subsection{Parametrization of the platelets}\label{sec:plateparam}

\begin{figure}[t]
  \begin{center}
  \begin{tabular}{c c}
	\includegraphics[width=0.45\textwidth]{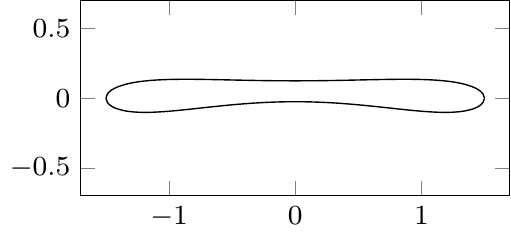}
    &
	\includegraphics[width=0.45\textwidth]{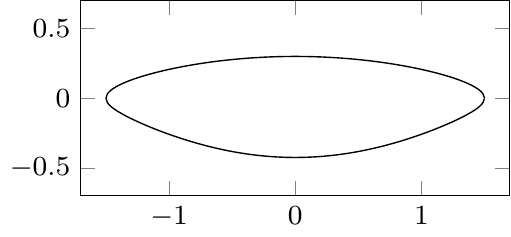}
    \\
    $\alpha_{top} = 0.5,\quad\alpha_{bot} = 0.1$ & $\alpha_{top} = 1.2,\quad\alpha_{bot} = 1.7$
    \\ \\
	\includegraphics[width=0.45\textwidth]{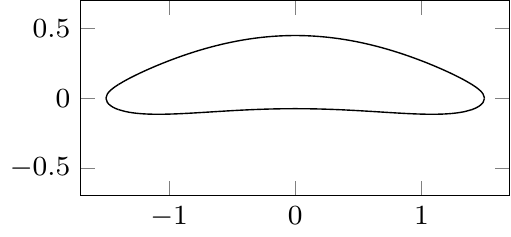}
    &
	\includegraphics[width=0.45\textwidth]{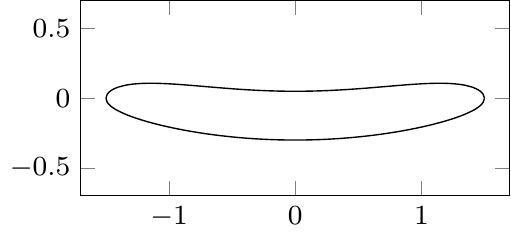}
    \\
    $\alpha_{top} = 1.8,\quad\alpha_{bot} = 0.3$ & $\alpha_{top} = 0.2,\quad\alpha_{bot} = 1.2$
  \end{tabular}
  \end{center}
  \caption{Prototypical templates for variations of the parameters
    $L_x,L_y,L_z$ (length) and $\alpha_{top},\alpha_{bot}$
    (shape). Within the blood flow, particles can appear at all angles of
    attack $\phi$. }  
  \label{fig:platelets}
\end{figure}

We model the platelets as variations of an ellipsoid with major axes $L_x\times L_y\times L_z$ with $L_x\approx  L_z\approx \unit[3]{\mu m}$ and $L_y\approx \unit[0.5]{\mu m}$. In $y$-direction upper $\alpha_{top}$ and lower $\alpha_{bot}$ semi-ellipsoids are modified to give them a more or less concave or convex shape. Alltogether, each particle is described by a set of 5 parameters $P=(L_x,L_y,L_z,\alpha_{top},\alpha_{bot})$. The surface of the platelets is given as zero contour of the levelset function
\[
\Phi\big(P;x,y,z\big)
=1 - R(P)^2
-\Big(\alpha+(1-\alpha)R(P)^2\Big)^{-2}
\left(\frac{2y}{L_y}\right)^2,
\]
where we define
\[
R(P)^2\coloneqq \left(\frac{2x}{L_x}\right)^2
+\left(\frac{2z}{L_z}\right)^2
\]
and
\[
\alpha\coloneqq \begin{cases}\alpha_{top} & y>=0\\
  \alpha_{bot} & y<0.
\end{cases}
\]
We assume that all parameters
$L_x,L_y,L_z,\alpha_{top},\alpha_{bot}$ are normally distributed with
means indicated above and with standard derivation $0.3$ for the
lengths $L_x,L_y,L_z$ and $0.4$ for the shape parameters
$\alpha_{top},\alpha_{bot}$. We drop particles that exceed the bounds
\begin{equation}\label{bounds}
  L_x,L_z\in [\unit[2.5]{\mu m},\unit[3.5]{\mu m}],\quad L_y\in  
  [\unit[0.15]{\mu m},\unit[1]{\mu m}],\quad
  \alpha_{top},\alpha_{bot}\in [0.2,2]. 
\end{equation}
In Fig.~\ref{fig:platelets} we show some typical shapes of the
platelets. 

Next, we indicate mass, center of mass and moment of inertia for a parametrized particle $P=(L_x,L_y,L_z,\alpha_{top},\alpha_{bot})$. Unless otherwise specified, all quantities are given in $\mu m$ and $g$. The mass of a particle $P$ is approximated by 
\begin{equation}\label{mass}
  m(P):=
  \rho_P L_xL_yL_z 0.2116
  \big(\sqrt{\alpha_{bot}+0.5313}
  +\sqrt{\alpha_{top}+0.5313}\big).
\end{equation}
This approximation is  based on a weighted one-point Gaussian quadrature rule. It is accurate with an error of at most  $2\%$ for all $\alpha\in [0.2,2]$. 

The center of mass for a particle $P$ is given by

\begin{equation}\label{centerofmass}
  m_x(P)=m_z(P)=0,\quad 
  m_y(P) = \frac{L_xL_y^2L_z\pi}{96m(P)}\big(\alpha_{top}-\alpha_{bot}\big). 
\end{equation}

The moment of inertia in the $x/y$ plane (the only axis of rotation that we will consider in the 2d simplification within this work) is given by
\begin{multline}\label{inertiaZ}
  I_z(P) = \rho_P\int_P \big(x^2+y^2)\,\text{d}(xyz)\\
  \approx
  \frac{L_xL_yL_z\rho_P\pi}{8}
  \Big(
  0.24+
  0.12(\alpha_{top}+\alpha_{bot})+0.0236(\alpha_{top}^2+\alpha_{bot}^2)\Big),
\end{multline}
which is accurate up to an error of at most $1\%$ for all $\alpha\in [0.2,2]$. These coefficients are computed once for each particle and stored as additional parameters. 

\paragraph{2d Simplification}

\begin{figure}[t]
  \begin{center}
    \includegraphics[width=0.8\textwidth]{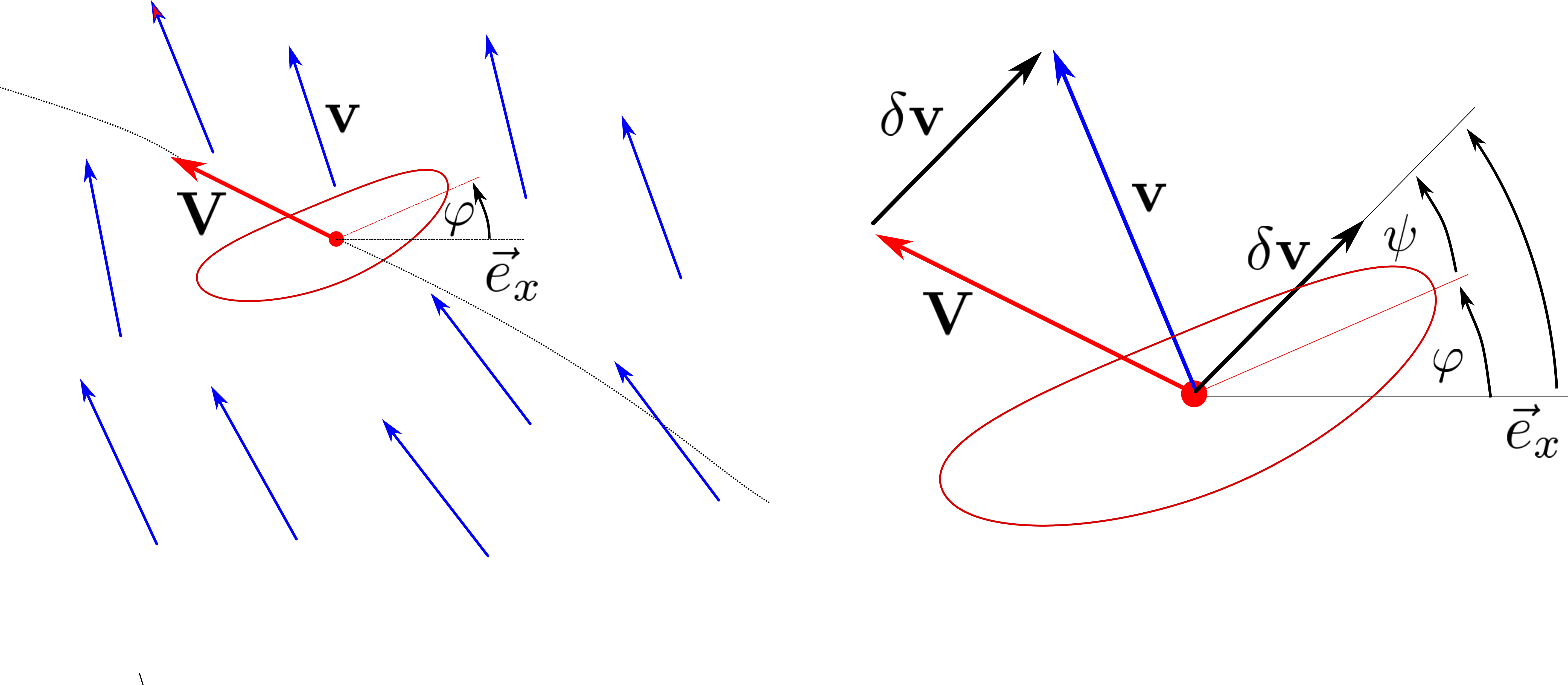}
    \caption{Left: typical configuration of one platelet (in red)
      within velocity field of the blood (blue arrows). The angle
      $\varphi$ is the orientation of the platelet relative to its
      standard orientation. 
      Right: the black arrow $\delta \vt=\vt-\vec V$ is the velocity
      acting on the platelet. By 
      $\psi\coloneqq \angle(\delta\vt,\vec e_x)-\varphi$ we denote the
      effective angle of attack.}
    \label{fig:angle}
  \end{center}
\end{figure}

To start with, we apply a two dimensional simplification of the problem by assuming that the blood vessel is a layer of infinite depth (in $z$-direction) and that it holds $\vt_3=0$ for the blood velocity and $\vec V_3=0$ for all particles. Further, given the symmetry of the particles w.r.t. rotation in the $x/y$-plane, no traction forces in $z$-direction will appear. Besides that, rotation is restricted to rotation around the $x/y$-plane. Hence, $\vec\Omega=(0,0,\omega)$ is described by a scalar component. A complete particle is then described by
\[
P = (L_x,L_y,L_z,\alpha_{top},\alpha_{bot};\vec X,\varphi,\vec V,\vec \Omega), 
\]
where $(L_x,L_y,L_z,\alpha_{top},\alpha_{bot})$ are the 5 shape parameters and, $\vec X$ is the (2d) position, $\varphi$ the orientation w.r.t. the $z$-axis, $\vec V$ the (2d) velocity and $\vec \Omega$ the angular velocity w.r.t. the $z$-axis rotation.

To describe the forces acting on the particle suspended in the Navier-Stokes fluid we denote by $\delta\vt\coloneqq\vt-\vec V$ the effective velocity vector, i.e. the relative velocity that is acting on the platelets. By $\psi\coloneqq\angle(\vec e_x,\vt-\Vt)-\varphi$ we denote the effective angle of attack which is the angle between relative velocity $\delta\vt$ and the current orientation of the platelet, see Fig.~\ref{fig:angle} (right) and~(\ref{forces:coefficients}). It is computed as 
\begin{equation}\label{angle}
  \psi\coloneqq\angle(\vec e_x,\vt-\Vt)-\varphi.
\end{equation}

\begin{figure}[t]
  \begin{center}
    \begin{minipage}{0.4\textwidth}
    \includegraphics[width=\textwidth]{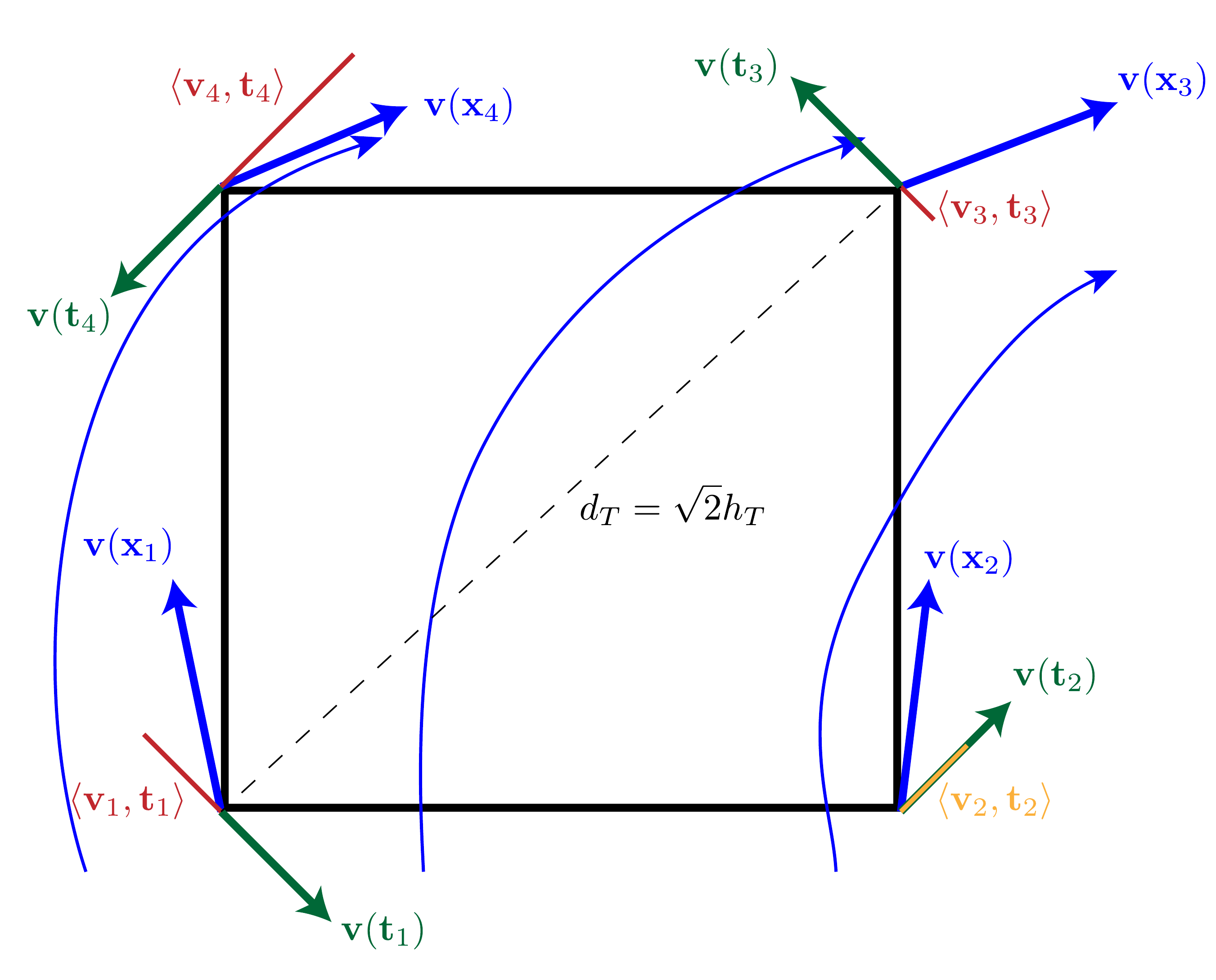}
    \end{minipage}\hspace{0.05\textwidth}
    \begin{minipage}{0.5\textwidth}
      We denote by $\vt(\xt_i)$ the Navier-Stokes velocity at node $\xt_i$, by $\vt(\mathbf{t}_i)$ the unit-tangential vector at this node in counter-clockwise orientation. With $\langle \vt_i,\mathbf{t}_i\rangle$ we denote the velocity contribution in tangential direction. Positive values are indicated in orange (at node $\xt_2$), negative contributions are in red.
    \end{minipage}
  \end{center}
  \caption{Approximation of the rotational velocity on an element $T$. }
  \label{fig:rotvelocity}
\end{figure}

Furthermore, denote by $\delta\omega\coloneqq\omegat(\vt)-\omega$ the relative angular velocity. The angular part of the Navier-Stokes velocity is locally reconstructed from the velocity field in every lattice, i.e. in every finite element cell $T$, by means of
\begin{equation}\label{rotvelocity}
  \omegat_T = \frac{1}{2d_T}\sum_{i=1}^4 \langle \vt(\xt_i),\vec t_i\rangle_2, 
\end{equation}
where $\xt_i$ and $\vec t_i$, for $i=1,2,3,4$ are the four nodes and tangential vectors of the lattice and $d_T=\sqrt{2}h_T$ is the diameter of the lattice, see also Figure~\ref{fig:rotvelocity}.

\subsection{Design of the artificial neural network}

\begin{figure}[t]
  \begin{center}
  \begin{minipage}{0.4\textwidth}
    \includegraphics[width=\textwidth]{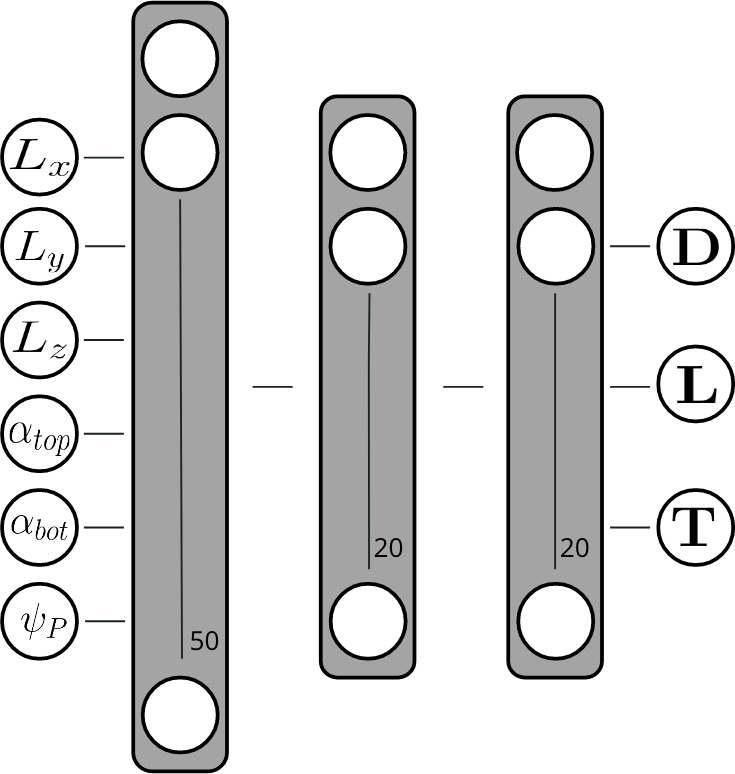}
  \end{minipage}\hspace{0.1\textwidth}
  \begin{minipage}{0.4\textwidth}
    \includegraphics[width=\textwidth]{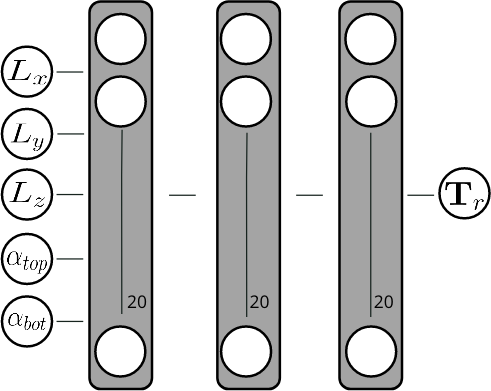}
  \end{minipage}
  \end{center}  
  \caption{Architecture for the neural networks ${\cal N}$ (left) for predicting drag, lift and pitching torque and ${\cal N}_r$ (right) for predicting the rotational torque. Both networks are fully connected feedforward networks with 3 hidden layers, having 50/20/20 and 20/20/20 neurons each.}
  \label{fig:network}
\end{figure}

We will train a deep neural network for determining the coefficients $C_d$, $C_l$, $C_p$ and $C_r$. We train two separate neural networks since the input data for $C_d$, $C_l$ and $C_p$ depends on the angle of attack $\psi$, while that of $C_r$ is invariant to the orientation of the particle. We call these artificial neural networks ${\cal N}$ and ${\cal N}_r$. Both take the platelet parameters  $(L_x,L_y,L_z,\alpha_{top},\alpha_{bot})$ as input. ${\cal N}$ further depends on the effective angle of attack $\psi_P$. Alltogether
\[
  {\cal N}:\mathds{R}^6\to \mathds{R}^3,\quad
  {\cal N}_r:\mathds{R}^5\to \mathds{R}.
\]
Both neural networks are fully connected feedforward networks with three hidden layers consisting of 50, 20 and 20 neurons in the case of the drag/lift network and 20 neurons each in the case of the rotational torque network. All neurons apart from the output layer are of ReLU type, i.e. using the activation function $f(x)=\max\{x,0\}$. 
Fig.~\ref{fig:network} shows the general configuration.

\subsection{Generation of the training data}\label{sec:train}

Training and test data is obtained by resolved simulations with random sampling of prototypical platelet shapes. Let $\Omega=\{\xt\in\mathds{R}\,:\, |\xt|_2<\unit[50]{\mu m}\}\setminus P$ be the open ball with radius $R=\unit[50]{\mu m}$ around a platelet $P$. Each platelet $P$ is constructed  by taking normally distributed values for $(L_x,L_y,L_z,\alpha_{top},\alpha_{bot})$ as indicated in Section~\ref{sec:plateparam}. Since only relative velocity and relative orientation matter, the platelets are fixed in the origin with $\vec V=0$ and $\Omega=0$ at angle $\varphi=0$.

For training, the Navier-Stokes equations are formulated in the units $\unit{\mu m}$ for length, $\unit{\mu m\cdot s^{-1}}$ for the velocity and $\unit{\mu g}$ for mass. With the blood viscosity $\mu = \unit[3]{\mu g\cdot \mu m^{-1}\cdot s^{-1}}$ the Stokes equations 
\begin{equation}\label{ns}
  -\mu \Delta \vt+\nabla p =0,\quad \div\,\vt=0,
\end{equation}
are considered.
We prescribe zero Dirichlet data on the platelet, $\vt=0$ on $\partial
P$, and set a freestream velocity on the outer boundary
$\partial\Omega\setminus\partial P$. This is either a uniform parallel
flow field or a uniform
rotational flow field that corresponds to the rotational velocity
$\omega = 2\pi$, both given as Dirichlet data  
\begin{equation}\label{dirichlet}
  \vt=0\text{ on }\partial P,\quad
  \vt^d_\psi\coloneqq\begin{pmatrix} 
  \cos(\psi) \\ \sin(\psi)\\ 0
  \end{pmatrix} \text{ or }
  \vt^r_\omega\coloneqq 2\pi \begin{pmatrix}
    -y\\ x \\ 0
  \end{pmatrix}\text{ on }\partial\Omega\setminus \partial P,
\end{equation}
where $\psi\in [0,2\pi]$ is the relative angle of attack. For $\vt^d$
it holds $|\vt^d_\psi|=\unit[1]{\mu m\cdot s^{-1}}$ and in case of the
rotational flow it holds $|\vt^r_\omega|=\unit[2\pi R]{\mu m\cdot
  s^{-1}}$, such that 
it  corresponds to a angular velocity of magnitude 
$2\pi$ in counter-clockwise direction around the $z$-axis.

The training data is generated as follows
\begin{algo}[Generation of the training data]\label{algotrainingdata}
  Let $N\in\mathds{N}$ be a prescribed number of experiments. For
  $n=1,2,\dots,N$
  \begin{enumerate}
  \item Generate a random particle $P_n$ that satisfies the
    bounds~(\ref{bounds})
  \item For four random angles of attack $\psi_{n,i}\in [0,2\pi]$,
    $i=1,2,3,4$, solve the Stokes equations for the directional
    Dirichlet data $\vt^d_{\psi_{n,i}}$ and compute\footnote{By
      $(\xt-\xt_P)^\perp$ and $\vt^{d,\perp}_{\psi_{n,i}}$ we denote
      the counter-clockwise rotation of the corresponding vectors by
      $90^\circ$.} 
    \[
    \begin{aligned}
      \vec D_{n,i}&:=\int_{\partial  P} \Big(\mu\nabla \vt
      - pI\Big)\vec n\cdot \vt^d_{\psi_{n,i}}\,\text{d}s\\ 
      \vec L_{n,i}&:=\int_{\partial  P} \Big(\mu\nabla \vt
      - pI\Big)\vec n\cdot \vt^{d,\perp}_{\psi_{n,i}}\,\text{d}s\\ 
      \vec T_{n,i}&:=\int_{\partial P}
      (\xt-\xt_P)^\perp \cdot
      \Big(\mu\nabla \vt
      - pI\Big)\vec n\text{d} s\\
    \end{aligned}
    \]
  \item Solve the Stokes equations for the rotational Dirichlet data
    $\vt^d_\omega$ and compute the rotational torque
    \[
    \vec T_{n,r}:=\int_{\partial P}
    (\xt-\xt_P)^\perp \cdot
    \Big(\mu\nabla \vt
    - pI\Big)\vec n\text{d} s.
    \]
  \end{enumerate}
\end{algo}
Hereby, a set of $4N$ training data sets $(P_n,\psi_{n,i};\vec
F_{n,i},\vec T_{n,i})$ and $N$ data sets for the rotational
configuration $(P_n;\vec T_{n,r})$ are generated in an offline phase.
Two different networks will be used for these two different settings.

The domain $\Omega$ is meshed with hexahedral elements and the finite
element discretization is build on equal-order tri-quadratic finite
elements for velocity and pressure. The curved boundaries (both the
outer boundary and the platelet boundary) are approximated in an
isoparametric setup to avoid dominating geometry errors
see~\cite[Sec. 4.2.3]{Richter2017}.  A very coarse 
mesh with initially only 12 hexahedras is refined twice around the
platelet boundary once globally. The resulting discretization has
about $2\,000$ elements and $60\,000$ degrees of freedom. Details on the discretization are given in Section~\ref{sec:model}. The resulting (stationary) discrete finite element formulation is given by
\begin{equation}
  \begin{aligned}
    \rho\nu\big(\nabla\vt,\nabla\phi\big)-\big(p,\div\,\phi\big) &=0
    &\quad\forall \phi&\in [V_h]^2\\
    \big(\div\,\vt,\xi\big)+\sum_{T\in\Omega_h} \frac{h_T^2}{\nu}
    \big( \nabla (p-\pi_hp),\nabla (\xi -\pi_h \xi)\big)_T &=0&\quad\forall \xi&\in V_h.
  \end{aligned}
\end{equation}
For the Stokes equation no transport stabilization must be added. 

Given $\vt,p$ the resulting forces are computed by
\begin{equation}
  \vec F =
  \int_{\partial P}\big( \mu\nabla \vt-pI\big)\nt\,\text{d}o,\quad 
  T = \int_{\partial P}(\xt-\xt_P)^\perp \cdot \big(\mu\nabla\vt-p
  I\big)\nt\,\text{d}o.
\end{equation}
The units of $\vec F$ and $T$ are
\begin{equation}\label{forces:units}
  [\Ft] = \frac{\unit{\mu m\cdot\mu g}}{\unit{s^2}},\quad
  [T] = \frac{\unit{\mu m^2\cdot\mu g}}{\unit{s^2}}. 
\end{equation}

\begin{figure}[t]
  \begin{center}

    \includegraphics[width=0.32\textwidth]{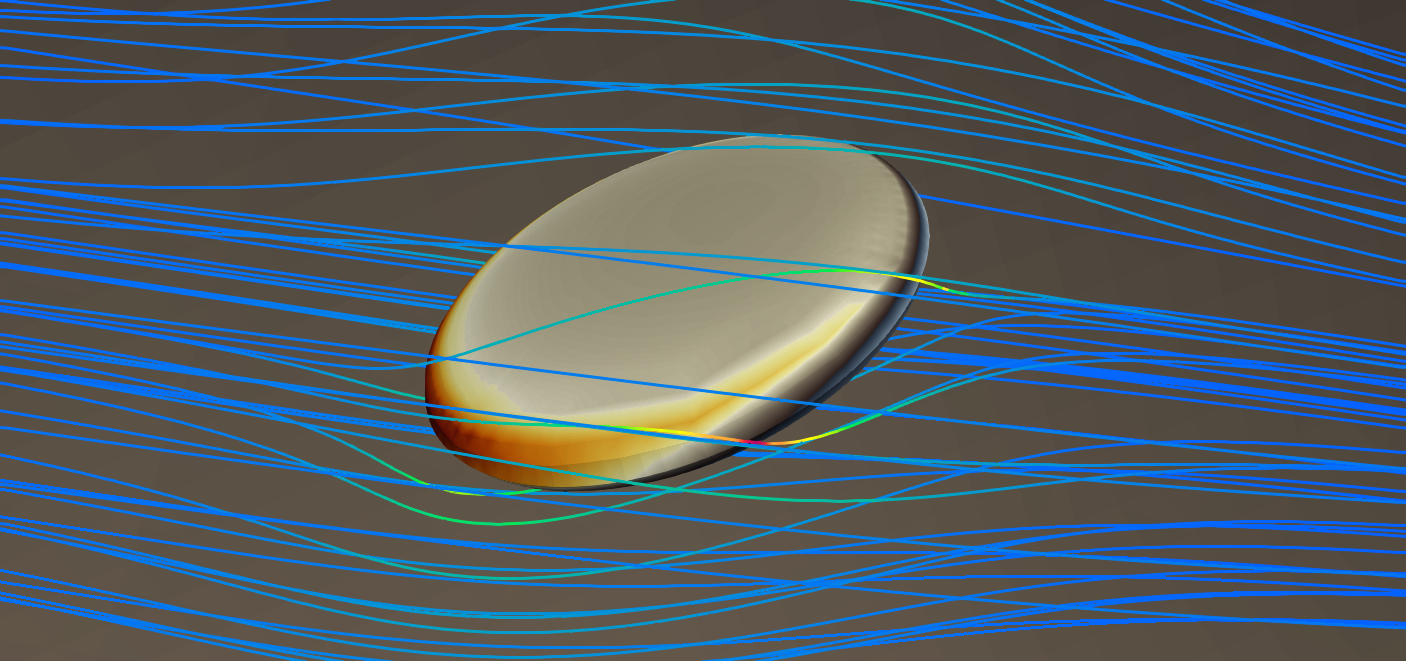}
    \includegraphics[width=0.32\textwidth]{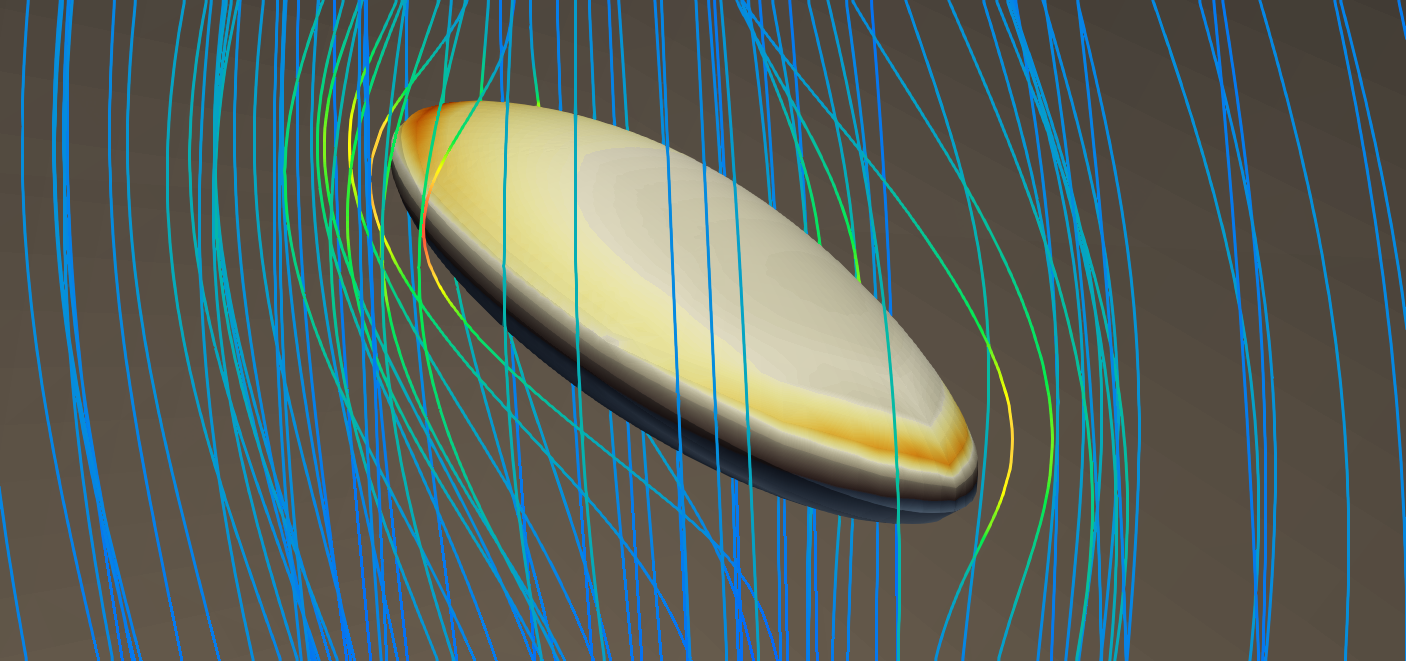}
    \includegraphics[width=0.32\textwidth]{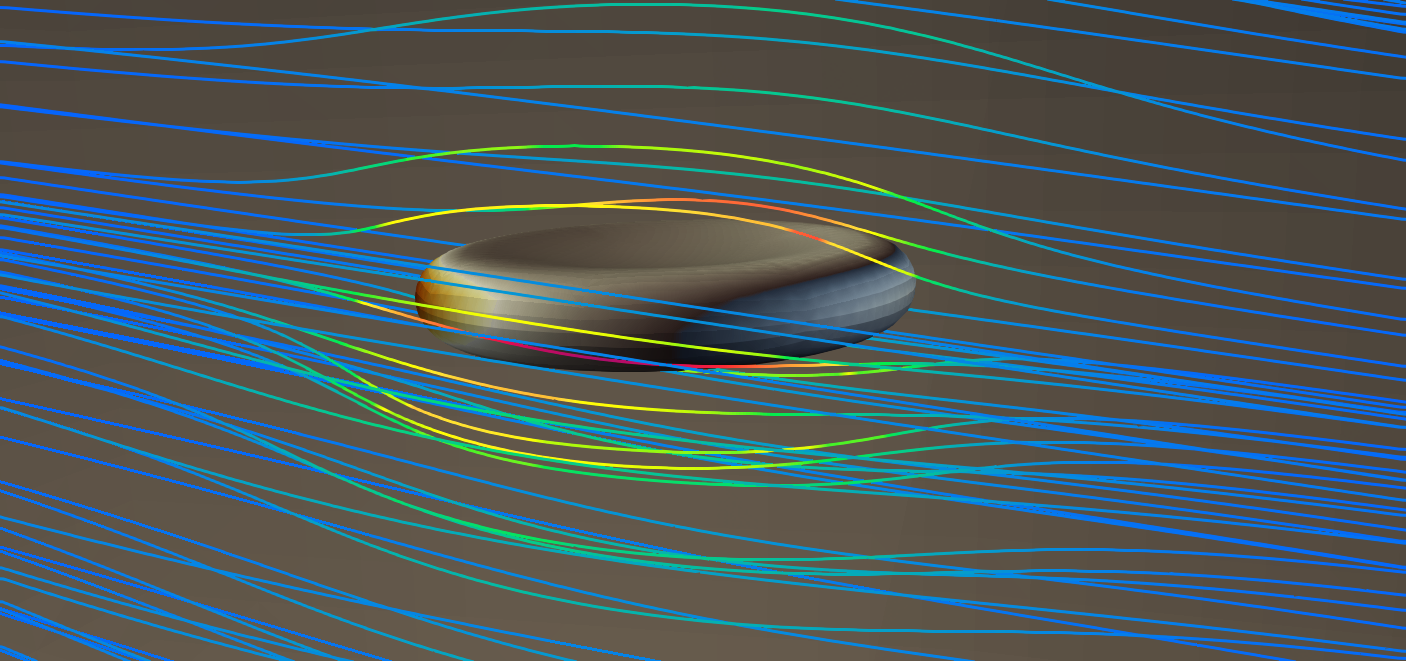}
    \includegraphics[width=0.32\textwidth]{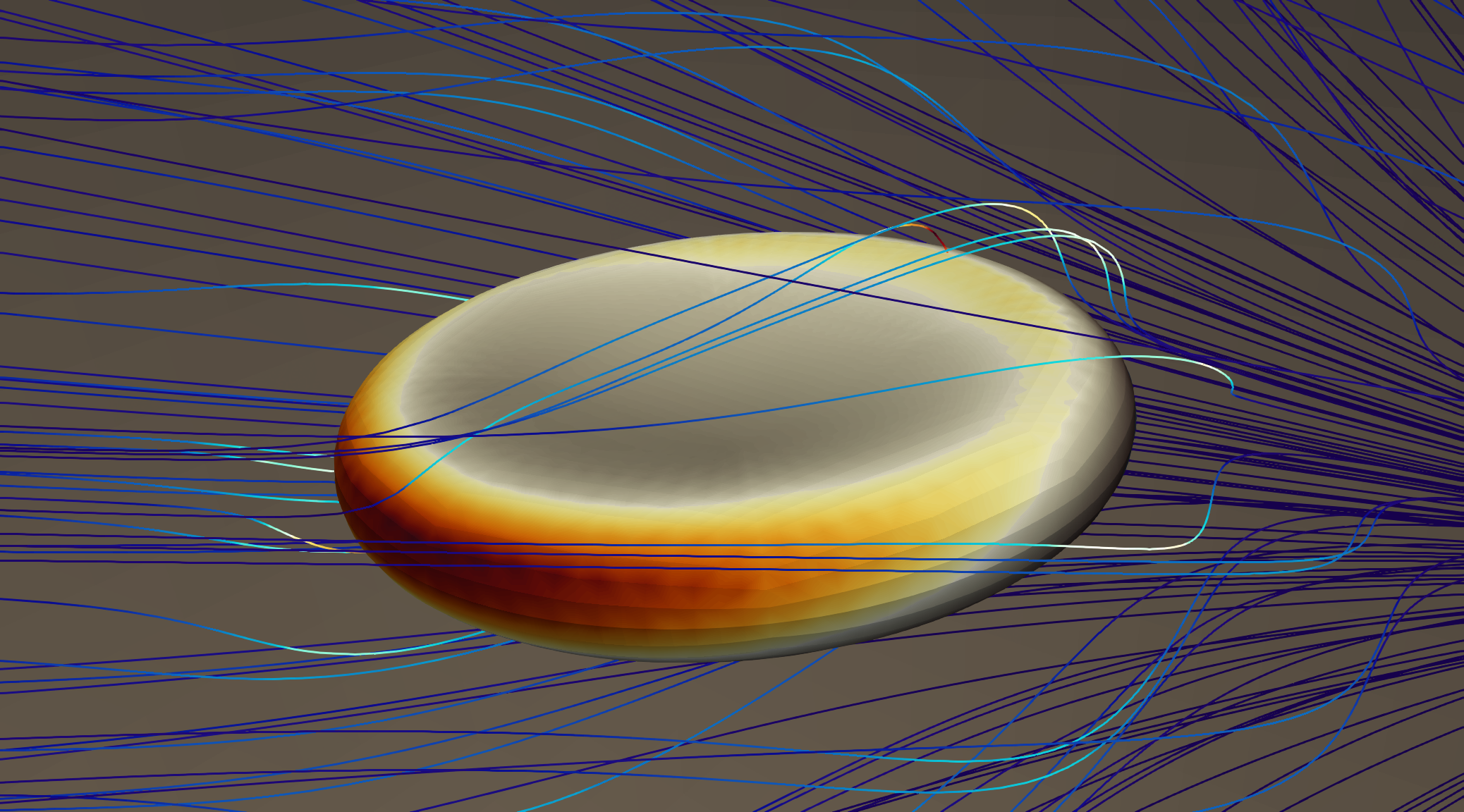}
    \includegraphics[width=0.32\textwidth]{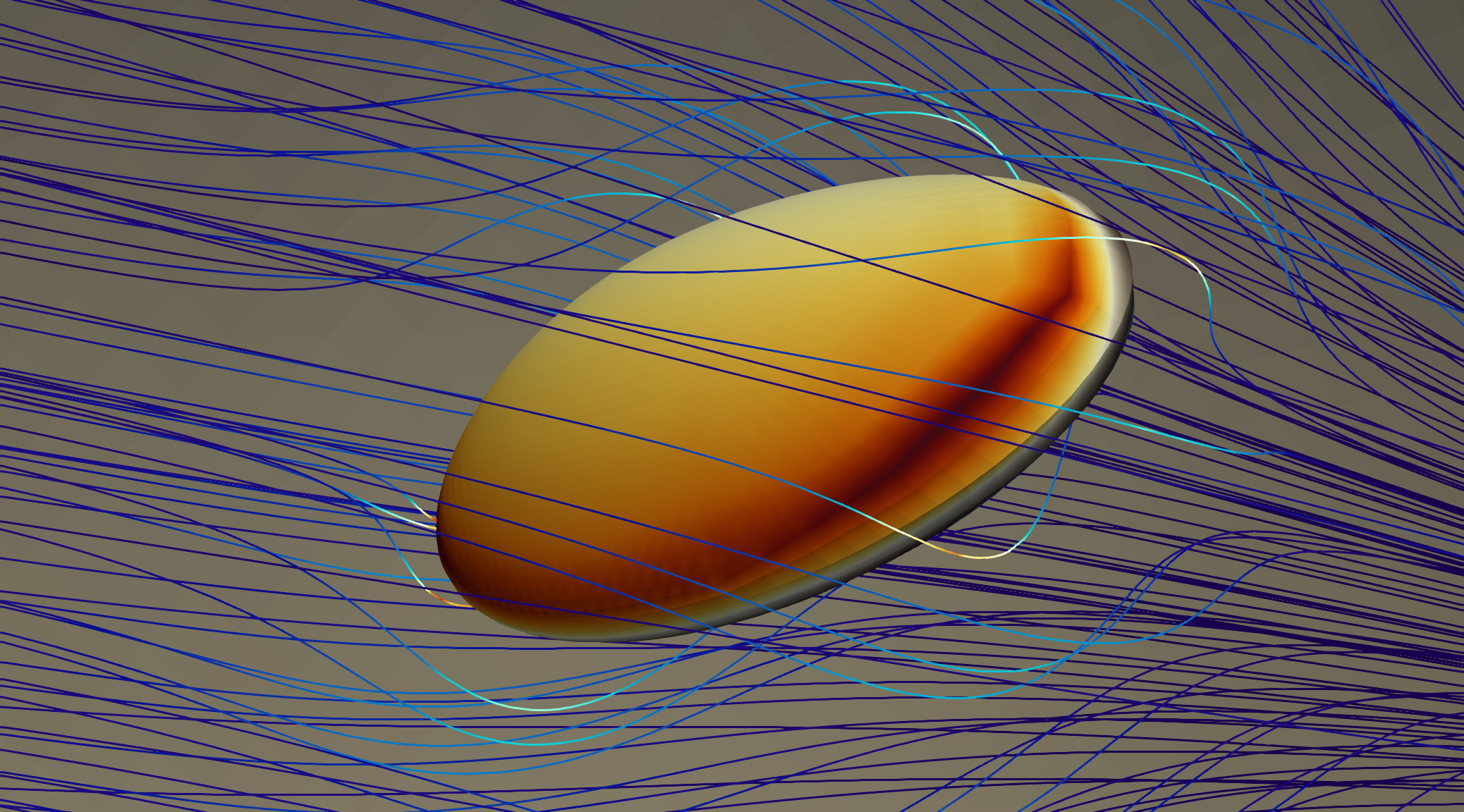}
    \includegraphics[width=0.32\textwidth]{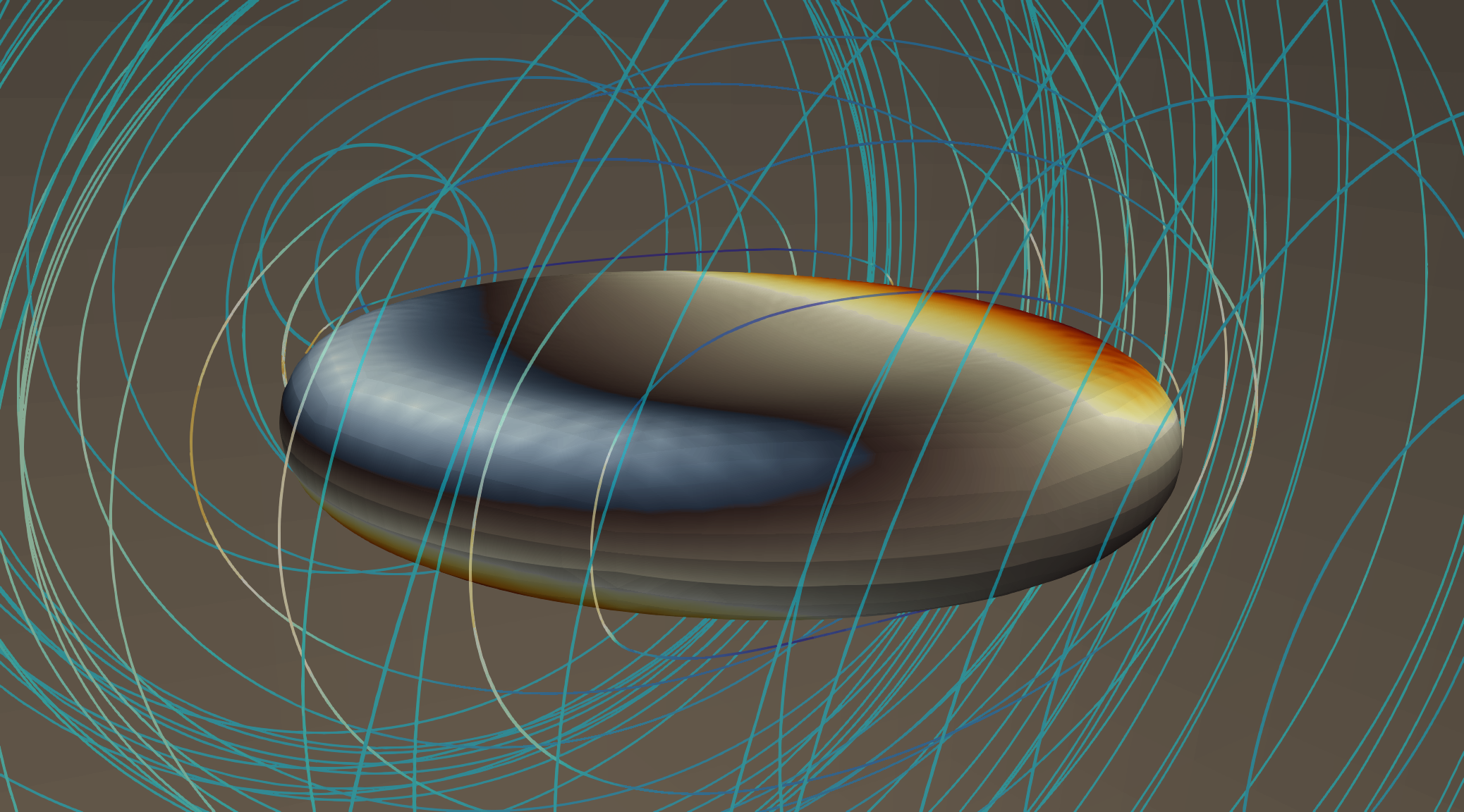}

  \end{center}
  \caption{Visualization of the resolved flow pattern around randomly
    created particles.   The platelets vary in size ($L_x\times
    L_y\times L_z$) and in their convexity, further we vary the angle
    of attack. The upper row shows three simulations with random particle using different angles of attack. In the lower row, we consider the same particle for each of the three simulations. The two figures on the left correspond to the directional inflow at different angles of attack while the plot on the right corresponds to a simulation with the rotational Dirichlet pattern on the outer boundary of the domain.}
  \label{fig:gasplate}
\end{figure}

Training and test data is computed on an Intel Xeon E5-2640 CPU at
2.40 GHz using 20 parallel threads. A total of $58\,500$ data sets
($46\,600$ for drag and lift, $11\,900$ for measuring the torque) have
been generated. The overall computational time for all these 3d
simulations was about 9 hours (less than one second for each
simulation). All computations are done in Gascoigne
3D~\cite{Gascoigne3D}. In Figure~\ref{fig:gasplate} we show snapshots of
three such simulations.

\subsubsection{Preparation and normalization of data / training of the neural network}

\begin{figure}[t]
  \includegraphics[width=\textwidth]{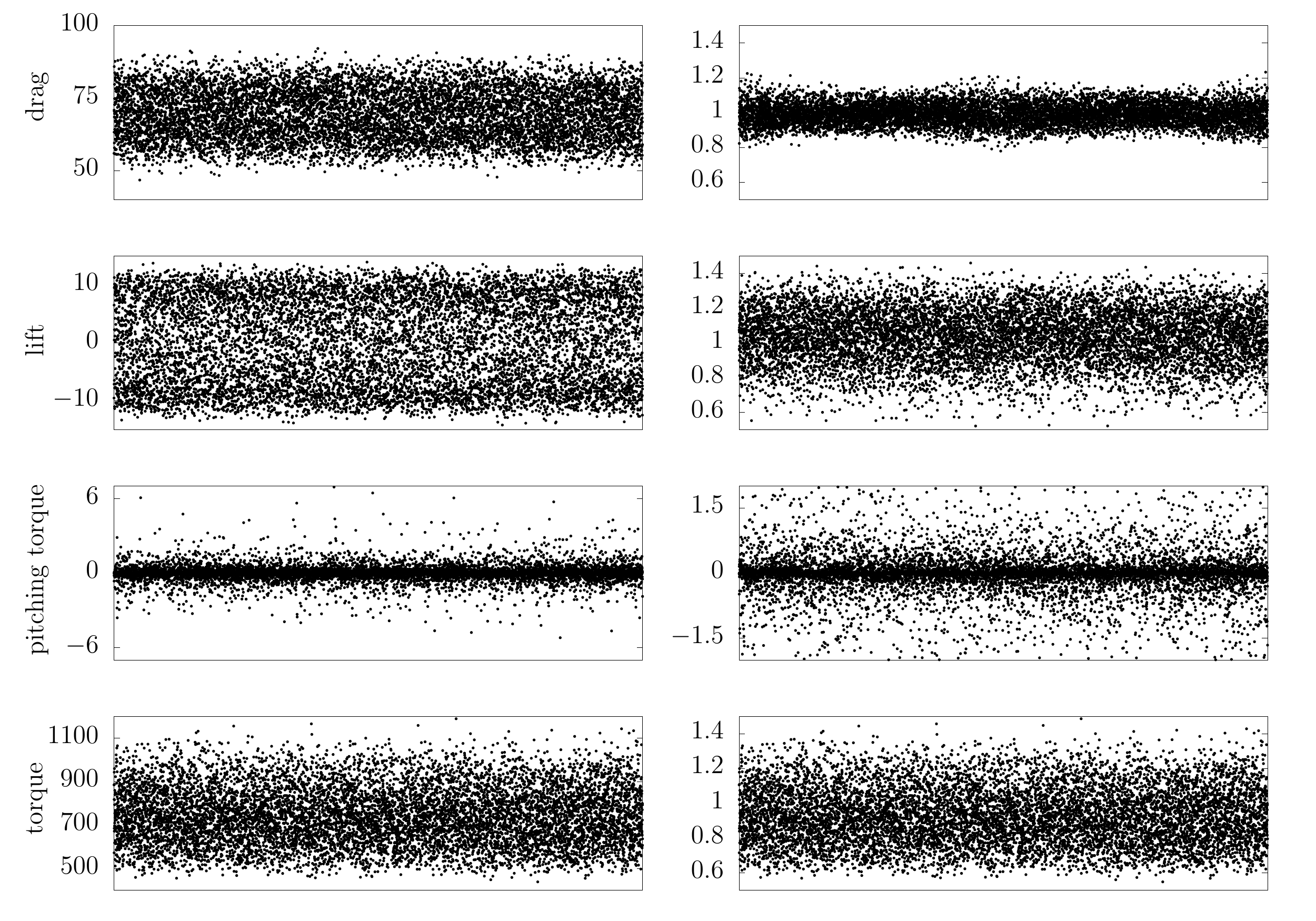}
  \caption{Visualization of the training data. Left: raw values coming
    from 5\,000 experiments (plotted along the x-axis) with randomly
    generated platelets and 
    random variation of the angle of attack. Right: scaled input data
    for the neural network according to~(\ref{scale}). By pre-scaling  
    the forces we can reduce the variation to about $20-40\%$. This
    remaining dependency of the quantities on the platelet- and
    flow-parameters will be learned in the artificial neural
    network.}
  \label{fig:scale}
\end{figure}

We produce a data set with $N$ entries with random particles. We start by 
extracting drag, lift and torque according to
Algorithm~\ref{algotrainingdata}. To
prepare the input data we encode as much model knowledge as
possible. Assume that $\vec D,\vec L,\vec T,\vec T_r$ are the vectors
containing drag, lift and pitching torque and rotational torque. Then, we define
the input data as (component-wise) 
\begin{equation}\label{scale}
  \Dt\coloneqq\frac{\vec D}{70-10\cos(2\varphi)},\quad
  \Lt\coloneqq\frac{\vec L}{10\sin(2\varphi)},\quad
  \Tt_p \coloneqq\frac{\vec T}{2\cos(\varphi)},\quad
  \Tt_r \coloneqq\frac{\vec T_r}{800}.
\end{equation}
These simple relations have been found manually by analyzing the
relation of the functional outputs on the different parameters. This
scaling reduces the variation of the forces over all experiments to
about than $10-40\%$ in the case of drag, lift and rotational
force. A rescaling of the pitching torque (which has a rather low
value) is more difficult since it depends on slight variations of the
particle symmetry.  

The neural network network is implemented in \emph{PyTorch}~\cite{NEURIPS2019_9015}, using the \emph{PyTorch C++
  API} which has been linked to our finite element framework Gascoigne 3D~\cite{Gascoigne3D}. The randomly generated data sets originating from the detailed Navier-Stokes simulations are split into $80\%$ serving as training data and the remaining $20\%$ as test data. As loss function we consider square $l_2$ norm of the error. The training of the two very small networks is accomplished in some few minutes. 

\subsubsection{Testing}\label{sec:test}

\begin{figure}[t]
  \begin{center}
    \includegraphics[width=\textwidth]{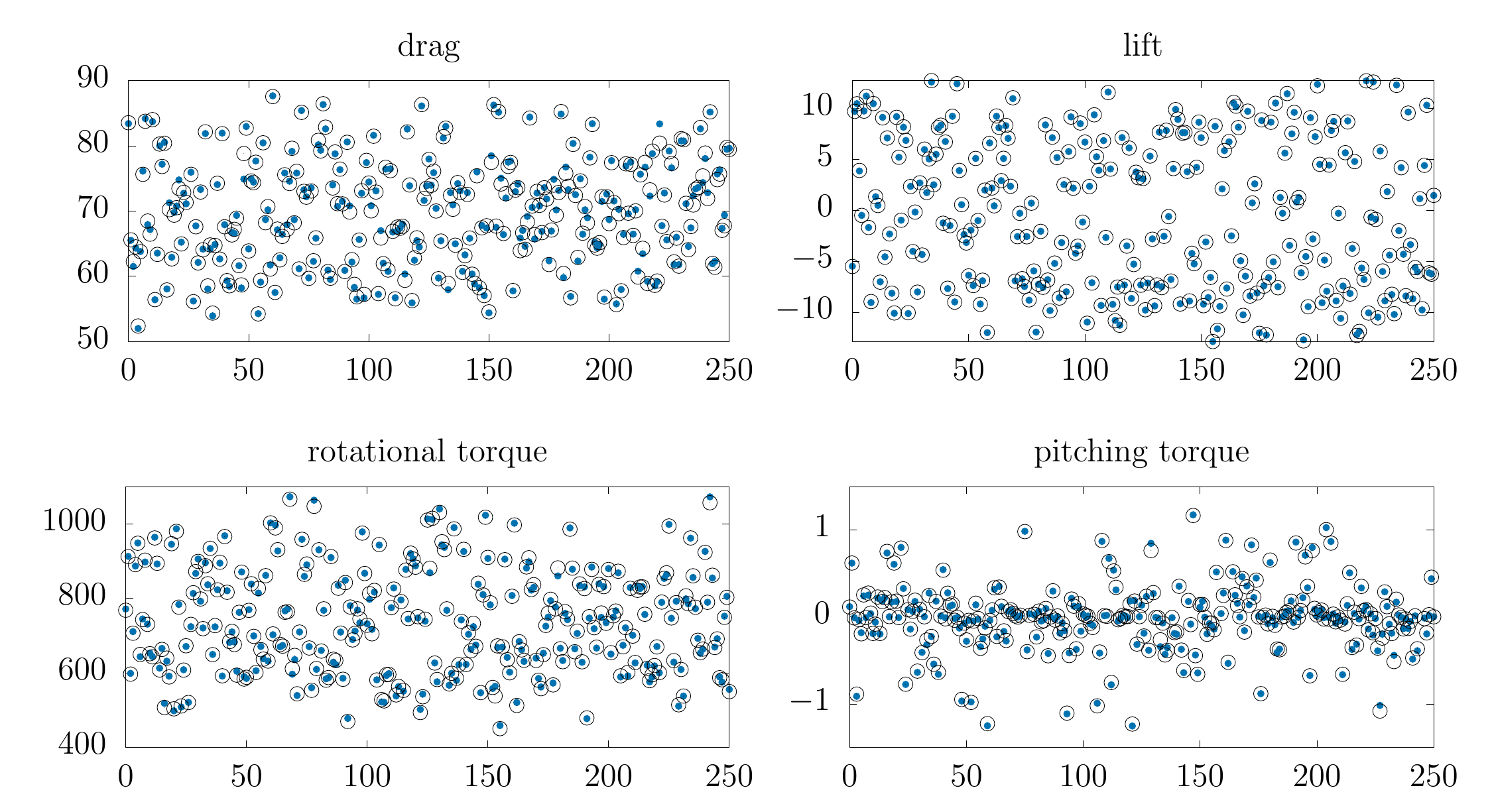}
    \caption{Performance of the neural network in predicting drag, lift and torque coefficients for the flow around randomly created platelets. For 250 random particles each (all have not been used in training the network) we compare the prediction (blue bullets) with the coefficients obtained in a resolved finite element simulation. The coefficients are given in the units of the training reference system described in~(\ref{forces:units}). }
    \label{fig:dragcirc}
  \end{center}
\end{figure}

To test the accuracy of the trained network we apply it to a set of testing data that was not used in the training of the network. In Figure~\ref{fig:dragcirc} we show for drag, lift and torque, 250  data points each that have been randomly taken from the test data set such that these data pairs have not been used in training. In the figure, we indicate the \emph{exact values} as taken from detailed finite element simulations that resolve the particle as large circles and the predicted DNN output as smaller bullets. We observe very good agreement in all three coefficients, best performance in the lift coefficient and highest deviation in the drag.

In Table~\ref{tab:test} we indicate the mean (measured in the $l_2$-norm) and the maximum error of the network applied to the training data and to the test data. Further, for getting an idea on the generalizability of the approach we also apply the network to additional testing data with random platelets, where at least one of the coefficients $(L_x,L_y,L_z,\alpha_{top},\alpha_{not})$ does not satisfy the bounds specified in~(\ref{bounds}). We note that such particles are not appearing in the coupled Navier-Stokes particle simulation framework.  The average errors appearing in all training and testing data is less than $1\%$. Maximum relative errors for few single particles reach values up to $4\%$ in the case of the pitching torque, which is most sensitive with values close to zero. Even if we consider data points that are not within the bounds, average errors are still small although substantial errors are found for single particles.

\begin{table}[t]
  \begin{center}
    \begin{tabular}{lcccccc}
      \toprule
      &\multicolumn{2}{c}{training data}
      &\multicolumn{2}{c}{test data}
      &\multicolumn{2}{c}{generalization}\\
      &avg&max&avg&max&avg&max\\
      \midrule
      Drag	& $0.40\%$ & $1.44\%$  & $0.40\%$ & $1.43\%$  & $1.35\%$ & $\phantom{1}6.30\%$  \\
      Lift	& $0.34\%$ & $1.32\%$  & $0.34\%$ & $1.33\%$  & $2.22\%$ & $19.90\%$  \\
      Pitching torque & $0.93\%$ & $3.75\%$  & $0.95\%$ & $3.87\%$  & $6.62\%$ & $55.79\%$  \\
      Rotational torque	& $0.45\%$ & $1.57\%$  & $0.45\%$ & $1.50\%$  & $2.21\%$ & $11.83\%$  \\
      \bottomrule
    \end{tabular}
    \caption{Accuracy of the neural network model for predicting
      drag, lift and pitching and rotational torque in percent. We
      indicate the values for the training 
      data, the test data and the hard test data that consists of data
      points outside the bounds~(\ref{bounds}).}
    \label{tab:test}
  \end{center}
\end{table}

\subsection{Application of the neural network}

The neural network predicts the coefficients for drag $\Dt$, lift
$\Lt$, pitching torque $\Tt_p$ and rotational torque $\Tt_r$, which
are scaled according to~(\ref{scale}). While drag, lift and pitching
torque depend on the effective angle of attack, the rotational torque
is a fixed value that must be predicted only once for each
particle. The former three values are recomputed whenever the
configuration is changing, i.e. before every advection step.

\begin{algo}[Neural network / particle / finite element coupling]\label{algo:loop}
Let $k$ be the macro time step used to predict the blood flow field,
$k_P\coloneqq k/M_p$  be the subcycling step for the particle dynamics and $N_p$
be the number of particles. For $n=1,2,\dots$ iterate
\begin{enumerate}
\item Solve the Navier-Stokes equations $t_{n-1}\mapsto t_n$
\item Transfer the local velocities from the finite element mesh to
  the particle lattice and locally compute the rotational velocity
  according to~(\ref{rotvelocity})
\item For $m=1,\dots,M_p$ subcycle the particle dynamics with step size $k_P$
  \begin{enumerate}
  \item For each particle $\{P_1,\dots,P_{N_p}\}$, compute the effective
    angle of attack  accorting to~(\ref{angle})
  \item Evaluate the deep neural network for all particles
    \[
    (L_x,L_y,L_z,\alpha_{top},\alpha_{bot},\psi)_i \mapsto
    (\Dt,\Lt,\Tt_p,\Tt_r)_i\quad i=1,\dots,N_p
    \]
  \item Rescale the coefficients according to~(\ref{scale}) and to
    correct the reference units\footnote{While the blood flow
      configuration is based in $\unit{mm}$ and $\unit{g}$, the particle
      configuration is set up in $\unit{\mu m}$ and $\unit{\mu g}$. }
    \[
    \begin{aligned}
      \vec D&=10^{-6}\cdot \big(70-10\cos(2\psi)\big) \Tt&\qquad
      \vec L&=10^{-6}\cdot 10\sin(2\psi) \Lt\\
      \vec T_p&=10^{-9}\cdot 2\cos(\psi)\Tt_p&
      \vec T_r&=10^{-9}\cdot 800 \Tt_r\\
    \end{aligned}
    \]
  \item Advect all particles and perform collisions according to Remark~\ref{rem:coll}.
  \end{enumerate}
\end{enumerate}
\end{algo}

\begin{remark}[Collisions]\label{rem:coll}
In order to detect and perform collisions we treat particles
as spheres with radius $L_x$ and use model described in~\cite{Luding1998}.
Since platelets constitutes only small part of the blood volume (less than 1\%)
collisions between them happen very rarely and this simplification does not
affect validity of presented approach.
\end{remark}
Usually we choose $M_p=100$ subcycling iterations within each macro step. For further acceleration, steps \emph{3.(a)-3.(c)} of the inner loop can be skipped in most of these inner iterations and it will be sufficient to recalculate the coupling coefficients in approximately every tenth step. 
\begin{remark}[Parallelization]
  Steps \emph{2., 3.(a)-3.(d)} are parallelized using OpenMP. Particles are organized on a lattice mesh. The dimensions of the lattice are generated such that a small number of particles reside in each lattice. This gives a natural way for parallelization and it also helps to keep the communication for performing particle-particle interactions local, compare~\cite{Luding2007} for details on this appraoch and for a review on further realization techniques.

  Step 4 of the algorithm involves the evaluation of the deep neural network. Here, we integrate C++ bindings of the library PyTorch~\cite{NEURIPS2019_9015} into Gascoigne~\cite{Gascoigne3D}. All particles are processed at once, such that the evaluation can be performed efficiently in the core of PyTorch. Considering larger networks or a larger number of particles, the use of a CUDA implementation is possible without further effort.

  Finally, step 1 of the algorithm requires to solve the Navier-Stokes equations in a finite element framework. The parallel framework that is used in Gascoigne is described in~\cite{FailerRichter2020,KimmritzRichter2010}.  
\end{remark}

\section{Numerical examples}\label{sec:numexamp}

\subsection{Evaluation of the Navier-Stokes / DNN particle coupling}\label{sec:results}

We study how the different shapes of the particles affect their movement and whether the neural network model is able to give distinguished responses for different particle types, even if the variations of the considered particles are small.
In order to do that we examine hydrodynamic forces acting on differently shaped particles. Simulations were performed for five particles (shown in Table~\ref{tab:partparam}) representing various shape features (symmetric, asymmetric, convex, convey and combinations).  

\begin{table}[h]
  \centering   
  \begin{tabular}{m{1.5cm}m{4cm}p{0.5cm}p{0.5cm}p{0.5cm}p{0.5cm}p{0.5cm}}
    \toprule
    \textbf{Particle}&\textbf{Shape}&
    \textbf{Lx}		& 
    \textbf{Ly}		& 
    \textbf{Lz}		& 
    \textbf{$\alpha_1$}	& 
    \textbf{$\alpha_2$}	\\
    \midrule
    1&
	\includegraphics[width=0.25\textwidth]{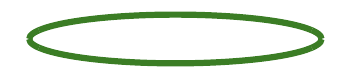}
	& 3.0 & 0.5 & 3.0 & 1.0 & 1.0\\
    2& 
    \includegraphics[width=0.25\textwidth]{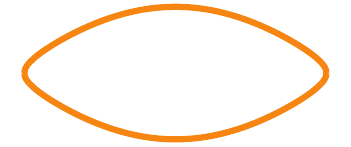} 
    & 3.1 & 0.8 & 2.5 & 1.7 & 1.7\\
    3& 		
    \includegraphics[width=0.25\textwidth]{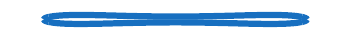} 
    &2.7 & 0.2 & 3.2 & 0.3 & 0.3 \\
    4& 		
	\includegraphics[width=0.25\textwidth]{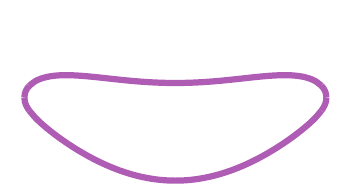} 
	& 3.1 & 0.7 & 2.8 & 0.3 & 1.7\\
    5& 		
	\includegraphics[width=0.25\textwidth]{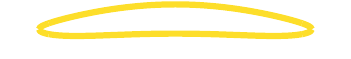}
	& 2.8 & 0.3 & 3.1 & 1.7 & 0.3\\
    \bottomrule
  \end{tabular}
  
  \caption{Parameters and shape of 5 test particles. The spatial dimensions are given in $\unit{\mu m}$.}
  \label{tab:partparam}
\end{table}

\begin{figure}[h]
	\begin{center}
		\includegraphics[width=\textwidth]{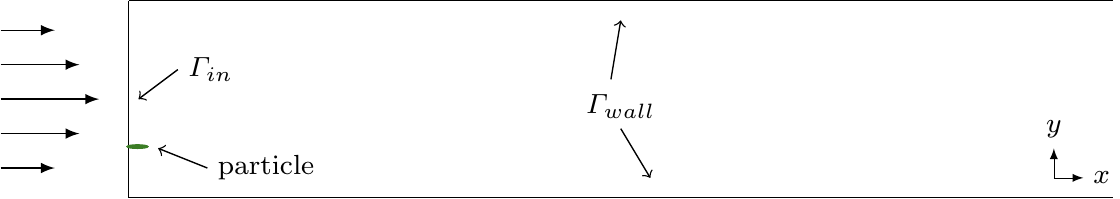}
		\caption{Spatial configuration of the considered model.}
		\label{fig:domain}
	\end{center}
\end{figure}

Our domain is a channel of diameter $L=\unit[2]{mm}$ and 
infinite length. The schematic geometry of the domain 
is described in Figure~\ref{fig:domain}. Platelets are 
variations of ellipsoids 
with major axes $L_x\times L_y\times L_z$ with 
$L_x\approx  L_z\approx \unit[3]{\mu m}$ and 
$L_y\approx \unit[0.5]{\mu m}$, for more details see 
Section~\ref{sec:plateparam}. 
The inflow data is defined by a time dependent parabolic 
inflow profile with inflow speed $\vt_{in}=\unit[5]{mm/s}$,
namely
\[
v_{in}(t) := y \frac{(2-y)}{2} t\cdot v_{in}.
\]
All five particles are initially located at  $y = 0.5176$, 
below the symmetry axis of the velocity profile such that 
a rotational velocity field attacks the particles. 
The fluid viscosity is set to $\mu=\unit[3]{mg/mm\cdot s}$, 
and particle and fluid density equal $\rho = \rho_p = 1.06\unit{mg/mm^3}$.
The parameters have been chosen so as to reflect a typical vessel, 
and realistic blood and platelet properties.

The simulations are carried out with the coupled interaction loop
described in Algorithm~\ref{algo:loop}. This means that after each
Navier-Stokes step, the fluid velocity is transferred to the particle
model and the coupling coefficients drag, lift and torques are updated
based on the previously trained neural network. Detailed simulations
around different particle shapes only enter the training phase by
generating random data sets. 

\subsubsection{Drag}\label{sec:drag}

\begin{figure}[h]
  \begin{center}
  \includegraphics[width=0.8\textwidth]{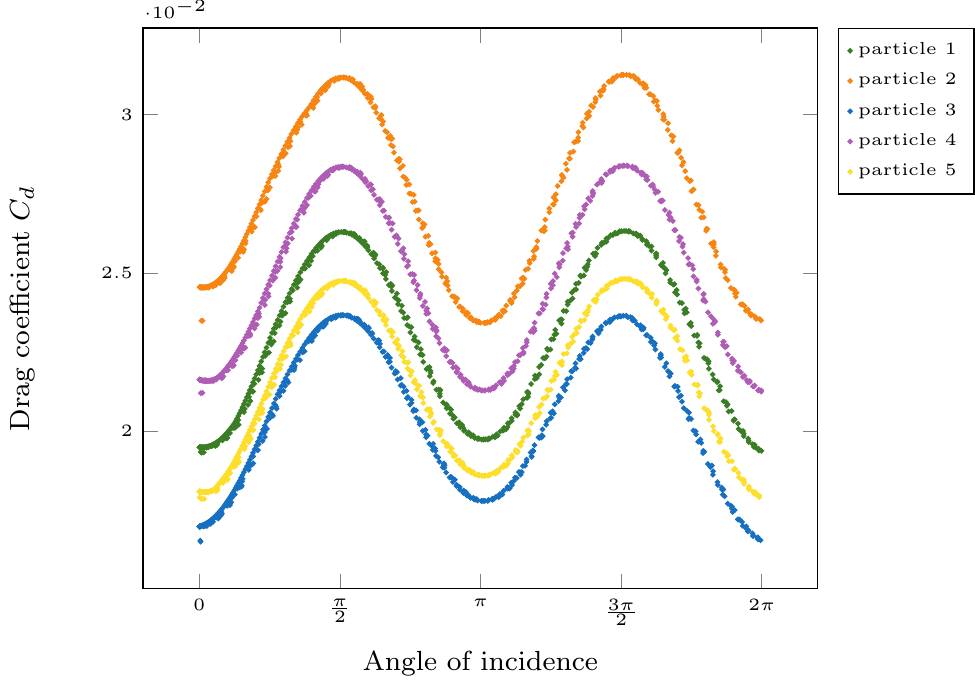}
  \end{center}
  \caption{Drag coefficient as a function of angle of incidence for the five particles defined in Table~\ref{tab:partparam}.}
  \label{fig:drag}
\end{figure}

Drag is a force acting opposite to the relative motion of the particle moving
with respect to a surrounding fluid. 
Shape-specific drag coefficients present in the literature are usually functions 
of the particle Reynolds number, angle of incidence and some shape parameters 
~\cite{Holzer2008,Rosendahl2010,Wachem2012}, while drag force itself usually depends 
on the properties of the fluid and on the size, shape, and speed of the particle.

In Figure~\ref{fig:drag} the drag coefficient is plotted as a function 
of the angle of incidence (effective angle of attack) for five considered 
particles. The first main observation lies in the increase of the drag 
values when the angle of incidence approaches $\psi=\frac{\pi}{2}$ 
and $\psi=\frac{3\pi}{2}$ so when particle is perpendicular
to the flow, which means the biggest cross sectional area
with respect to the flow. Correspondingly,
the drag decreases when the angle of incidence reaches $\phi=0$ or $\phi=2\pi$
so when the cross sectional area gets smaller.
Qualitatively, the present results are in good agreement with those issuing from 
the literature since a similar trend is observed (see e.g.~\cite{Sanjeevi2018}).

Furthermore, Figure~\ref{fig:drag} shows various drag coefficient values for 
different particles. Particle~2 is characterized by the highest value of the drag. 
Reasons may be threefold: big size of the particle in comparison to others 
and hence bigger cross sectional area and higher particle Reynolds number. 
In contrast, particle~3 is characterized by the lowest value of the drag, 
which is a result of its small size in comparison to other particles. 
Particle~1 is an ellipse and serves as a reference. Its drag coefficient 
is in the middle which is in the line with intuition - particle~1 has intermediate
values both in terms of size and convexity/concavity.

\subsubsection{Lift}\label{sec:lift}

\begin{figure}[h]
  \begin{center}
  \includegraphics[width=0.8\textwidth]{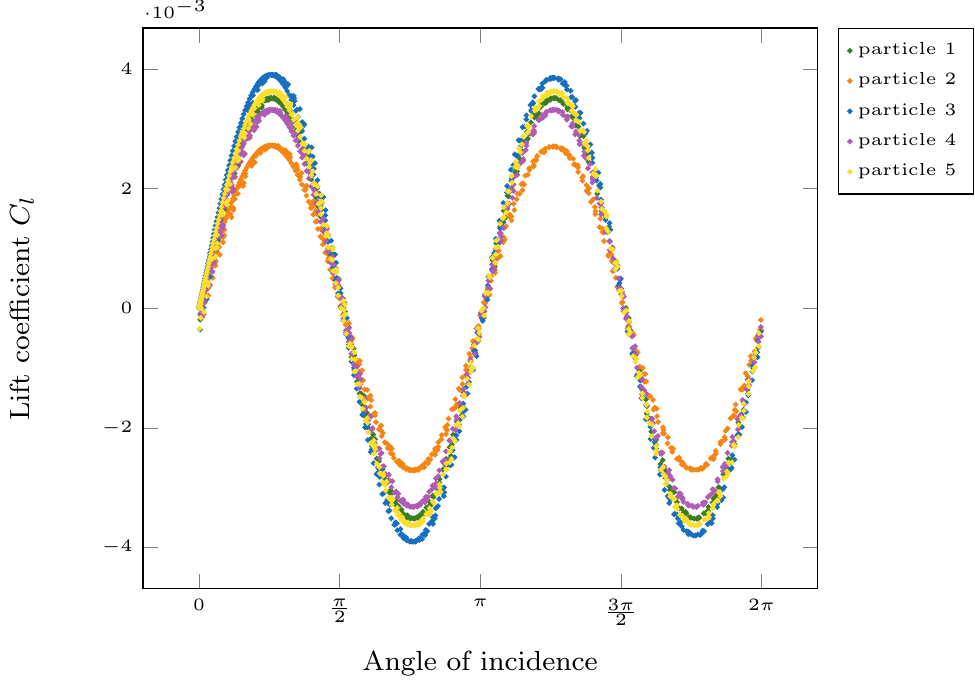}
  \end{center}
  \caption{Lift coefficient as a function of angle of incidence for the five particles defined in Table~\ref{tab:partparam}.}
  \label{fig:lift}
\end{figure}

Lift force on a particle is a result of non-axisymmetric flow field. 
The pressure distribution on the surface of a particle
inclined to the flow direction no longer follows the symmetry of that particle.
This gives rise to a lift force due to the displacement of the center of pressure. 
Lift acts in the direction perpendicular to the fluid velocity and is present when
the particles principle axis is inclined to the main flow direction. 
As in the case of drag, lift coefficient is usually a function 
of the particle Reynolds number, angle of incidence and some shape parameters 
~\cite{Holzer2009,Ouchene2016,Wachem2012}, while lift force itself usually depends 
on the properties of the fluid and on the size, shape, and speed of the particle.

The lift coefficient behaviour at various angles of incidence for five studied 
particles is presented in Figure~\ref{fig:lift}. The figure shows 
that the lift coefficient reaches its maximum when the angle of incidence
reaches $\psi=\frac{\pi}{4}$ or $\psi=\frac{5\pi}{4}$ and its minimum when  
$\psi=\frac{3\pi}{4}$ or $\psi=\frac{7\pi}{4}$ and is equal to $0$
for $\psi\in\{0,\frac{\pi}{2},\pi,\frac{3\pi}{2}\}$. These results
are consistent with the definition of the lift force and are 
similar to other studies (see e.g.~\cite{Ouchene2015,Sanjeevi2018}). 

Moreover, one can notice that the lift coefficient takes the lowest value 
for particle~3 and the highest value for particle~2. It results from the 
difference in surface area, 
which is small for particle~3 and big for particle~2. 
Similarly to the drag, the lift of the reference particle~1 is in the middle 
which also corresponds to the intermediate value of its surface area.

\subsubsection{Rotational torque}\label{sec:rottorq}

\begin{figure}[h]
  \begin{center}
  \includegraphics[width=0.8\textwidth]{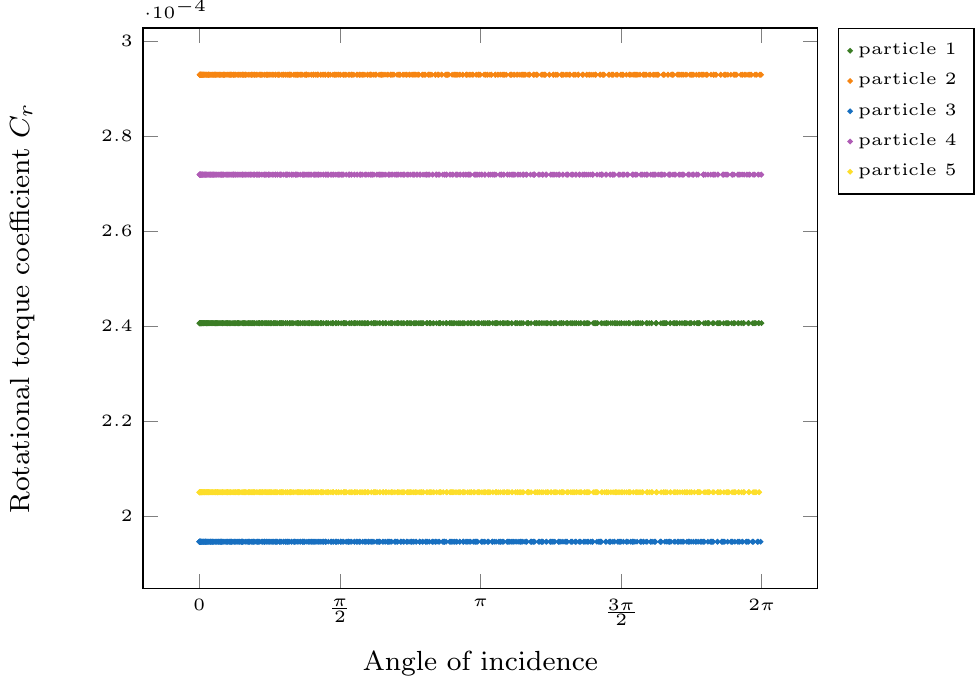}
  \end{center}
  \caption{Rotational torque coefficient as a function of angle of incidence for the five particles defined in Table~\ref{tab:partparam}. The rotational torque does not depend on the orientation of the particles since it is triggered by the symmetric rotational flow around the particle.}
  \label{fig:rottorq}
  
\end{figure}

There are two contributions to the rotational motion of the particle. 
The first is the inherent fluid vorticity, which acts on the particle 
as a torque due to the resistance on a rotating body. 
%Rotational torque originates from the relative rotation of the particle with respect to the fluid. 
%torque due to the viscous forces at the particle surface

Figure~\ref{fig:rottorq} illustrates the rotational torque coefficient plotted 
as a function of the angle of incidence for five examined particles.
One can notice that the magnitude of the coefficients corresponds 
to the surface area of particle, with particle~2's rotational 
torque coefficient being the highest, while particle~3 experiencing 
the smallest rotational torque. 

These results are consistent with the definition of the rotational torque
and qualitatively are similar to those obtained in the literature
(see e.g. \cite{Wachem2012}).

\subsubsection{Pitching torque}\label{sec:pitchtorq}

\begin{figure}[h]
  \begin{center}
  \includegraphics[width=0.8\textwidth]{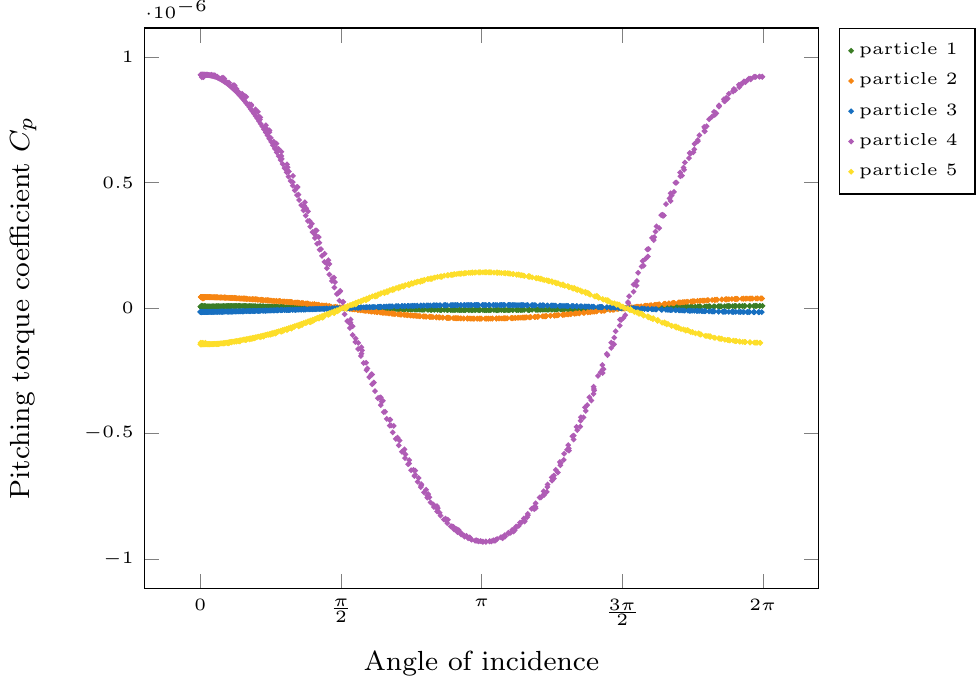}
  \end{center}
  \caption{Pitching torque coefficient as a function of angle of incidence for the five particles defined in Table~\ref{tab:partparam}.}
  \label{fig:pitchtorq}
\end{figure}

Since the center of pressure of the total aerodynamic force acting on each particle does not coincide with the particle's center of mass, a pitching torque is generated. This is the second factor that contributes to rotational motion. It accounts for the periodic rotation of the particle around an axis parallel to the flow direction.
%and attempts to increase the particle’'s angle of incidence.

\begin{figure}[h]
  \begin{center}
    \begin{tabular}{c c}
	\includegraphics[width=0.4\textwidth]{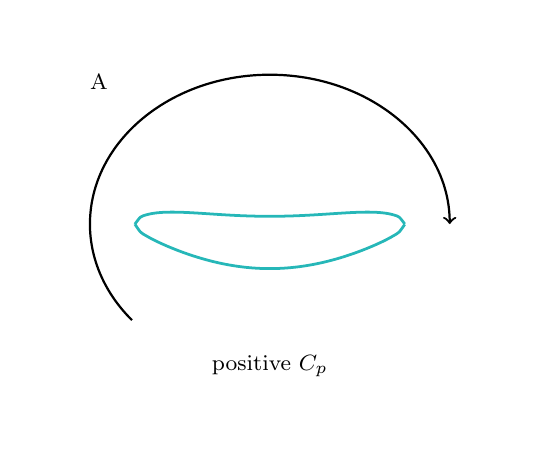}
      &		
	\includegraphics[width=0.4\textwidth]{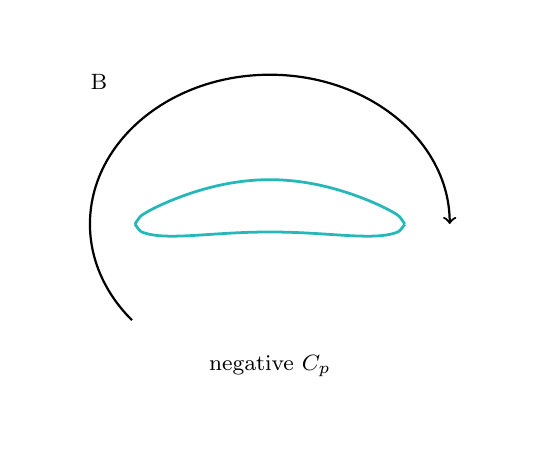}
    \end{tabular}
  \end{center}
  \caption{The pitching torque coefficient $C_p$ depends on particle settings.}
  \label{fig:pitchmode}
\end{figure}

In Figure~\ref{fig:pitchtorq} the pitching torque coefficient is plotted 
as a function of the angle of incidence for all five considered particles.
One can notice that the pitching torque coefficient is equal to $0$
for $\psi=\frac{\pi}{2}$ and $\psi=\frac{3\pi}{2}$ for all five particles,
so when particles are perpendicular to the flow. It means that for asymmetric particles,
i.e. particle~4 and 5, the pitching torque is $0$ when they are set symmetrically
with respect to the flow. It may imply that this is their preferred orientation. 
In case of particle~4 the pitching torque coefficient reaches its minimum
when the angle of incidence is $\psi=\pi$ and its maximum when 
$\psi=0$ or $\psi=2\pi$ are reached. It is caused by its asymmetric shape and setting 
with respect to the direction of the local fluid vorticity, 
namely particle~4 is convex \say{at the bottom} and concave 
\say{at the top} (for angle of incidence $\psi=0$ or $\psi=2\pi$), 
while the fluid around it is moving clockwise (see Figure~\ref{fig:pitchmode}). 
For particle~5 the situation is analogous, however
it is convex \say{at the bottom} and concave \say{at the top}. 
This is consistent with what happens for particle~4 and is reflected on the plot. 
For the remaining particles 1, 2, 3, the pitching torque is equal or close to $0$, 
which results from their symmetry.  

In the case of the pitching torque coefficient it is not straightforward 
to make a comparison between presented trends and those obtained 
in literature (e.g.~\cite{Holzer2009,Ouchene2016,Wachem2012}).
Most simulations are performed for non-spherical but symmetric particles.
Therefore, the discrepancy cannot be easily explained.

\subsubsection{Oscillatory translational motion}\label{sec:oscmot}

\begin{figure}[!]
  \begin{center}
  \includegraphics[width=0.8\textwidth]{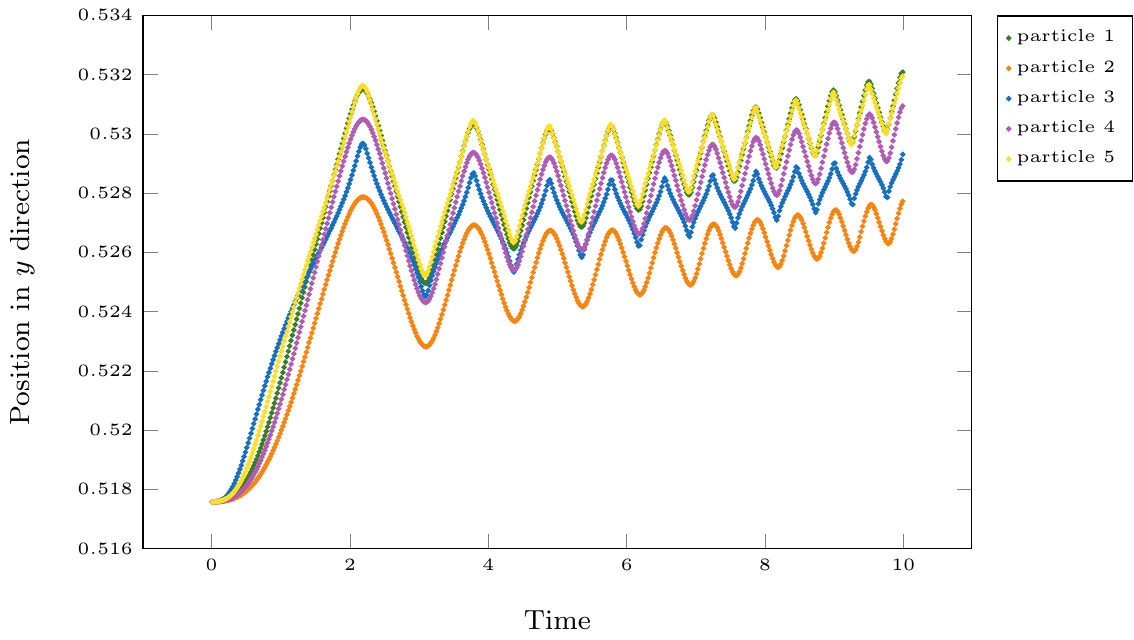}
  \end{center}
  \caption{Position in $y$ direction for the five particles defined in Table~\ref{tab:partparam}.}
  \label{fig:y}
\end{figure}

\begin{figure}[h]
	\begin{center}
	\includegraphics[width=0.8\textwidth]{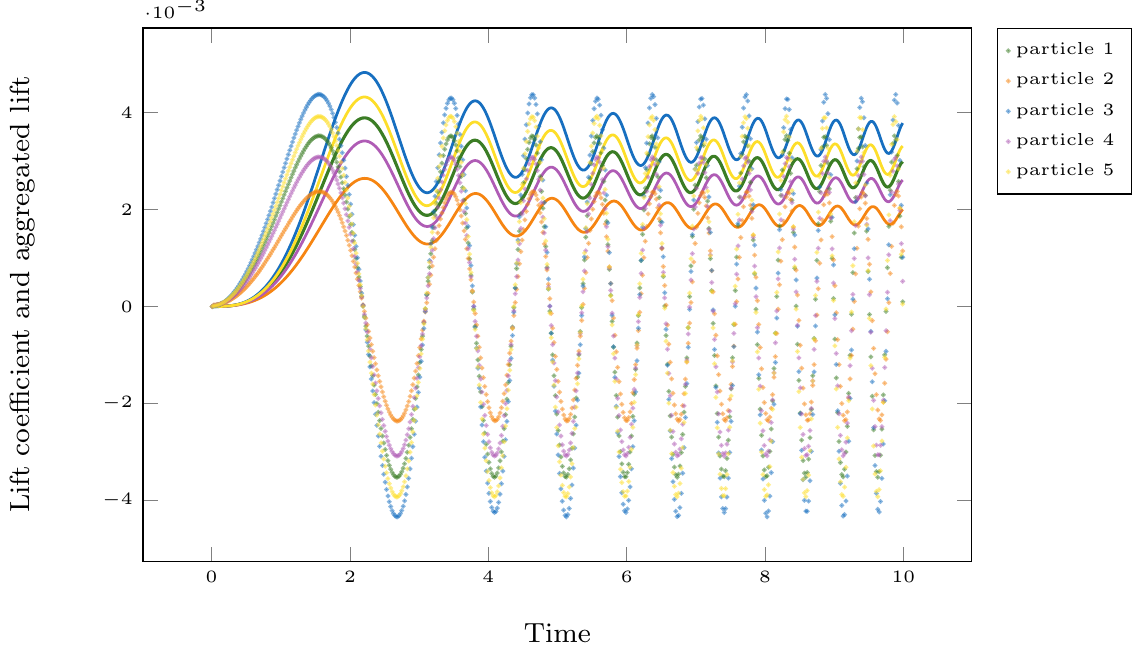}
	\end{center}
	\caption{Comparison of the lift coefficient for the different
          particles. In bold lines we show the aggregated lift over time.}
	\label{fig:liftint}
\end{figure}

\begin{figure}[h]
  \begin{center}
 \includegraphics[width=0.8\textwidth]{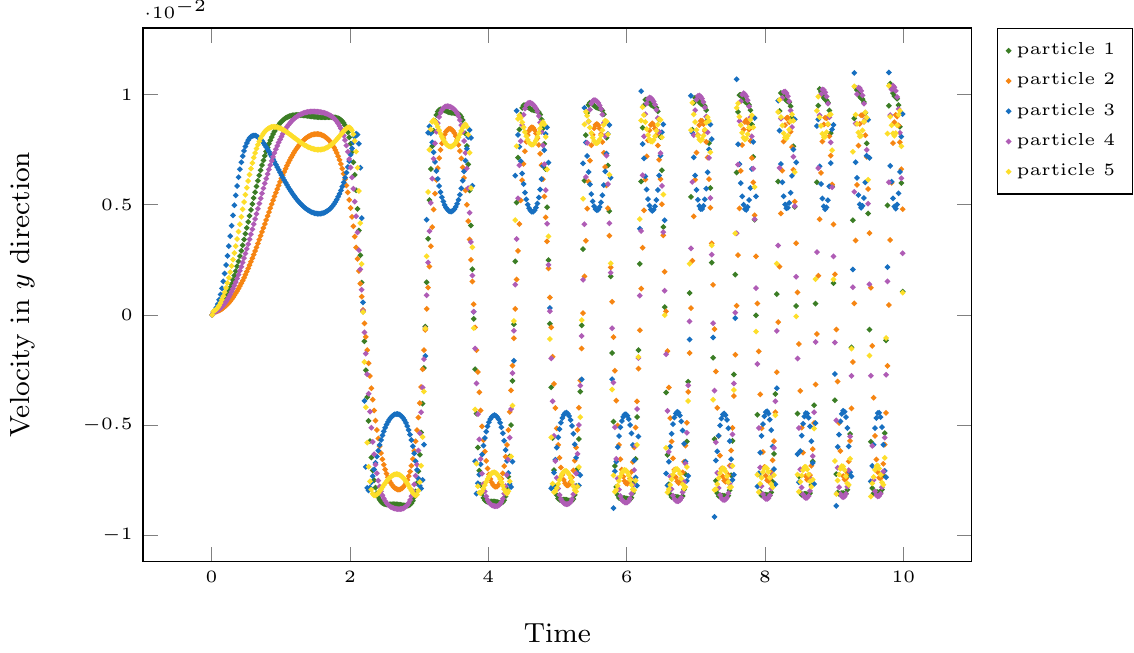}
  \end{center}
  \caption{Velocity in $y$ direction for the five particles defined in Table~\ref{tab:partparam}.}
  \label{fig:vy}
\end{figure}

The translational motion of non-spherical particles is characterized 
by an oscillatory motion. This is due to the fact that
the pressure distribution causes the hydrodynamic forces to work 
at the center of pressure rather than at the center of mass. 
The non-coincidence of the center of pressure and center of mass 
causes the sustained oscillations (see Figure~\ref{fig:y}).
Moreover, it is observed that every particle is also slowly 
moving up towards the horizontal axis of symmetry of the domain 
(all particles start below the axis, see Figure~\ref{fig:domain}).
In Figure~\ref{fig:liftint} evolution in time of the aggregated 
lift of five studied particles is plotted together with the
evolution in time of their lift coefficients. 
One can easily see that the oscillatory motion shown 
in Figure~\ref{fig:y} is a direct consequence of the lift force 
acting on the particles, while upward motion
results from the aggregated lift being positive all the time. 
The behaviour of the $y$-velocity is also worth noting 
(see Figure~\ref{fig:vy}). One can notice
that some particles (i.e. 1, 3, 5) decelerate when they are reaching
local minimum or maximum.   
Those local maxima and minima appear for angle of incidence 
$\psi\in\{\frac{\pi}{4},\frac{3\pi}{4},\frac{5\pi}{4},\frac{7\pi}{4}\}$,
so when particles are inclined to the flow direction. 
Particle~3, the thinnest one, is subjected to the highest deceleration, 
whereas particles~2 and~4 move more smoothly.

\subsection{Performance of the coupled model for many particles}\label{sec:sim}

In this section, we demonstrate the efficiency 
of the finite element/neural network approach and present
numerical results with a multitude of particles. We test
the computational effort for the particle model in comparison to the
finite element Navier-Stokes discretization. Although some effort has
been spent on the multicore implementation based on
OpenMP, our implementation is by no means a high
performance code. In particular, no use of GPU acceleration within the
particle model is applied, neither in the coupling to the neural
network model nor in the particle dynamics itself. Both is possible
and in parts already standard in available software packages such like
PyTorch C++\cite{NEURIPS2019_9015} or particle dynamics libraries such
as LAMMPS~\cite{LAMMPS}.

All computations have been carried out on a two-socket system with
Intel Xeon E5-2699A v4 processors running at 2.40 Ghz.

We will describe a prototypical blood-flow configuration and discuss
the scaling of the implementation with respect to the number of
cores. In particular we will investigate the relation between
computational effort used in the particle model and in the
Navier-Stokes solver. As in Section~\ref{sec:results} 
all the parameters have been chosen 
so as to resemble vessel, blood and platelet properties.

\subsubsection{Parallelization}

The finite element model is implemented in Gascoigne
3D~\cite{Gascoigne3D} and outlined in Section~\ref{sec:fluid}, the
discrete systems are approximated with a Newton-Krylow solver using
geometric multigrid preconditioning. Basic finite element routines
and the linear algebra workflow is partially parallelized using
OpenMP, see~\cite{FailerRichter2020} for details. Since the mesh
handling and i/o are not parallelized, substantial speedups are only
reached for complex 3d problems. 

The particle model is based on a regular lattice mesh that covers the
computational domain. The lattice elements of size $L_h\times L_h$
contain the individual particles. Detection of particle-particle
collision is limited to those particles that reside in the same
lattice element or that belong to directly adjacent elements. This
substantially helps to reduce the computational effort which scales
quadratically with the number of particles within each lattice
element. Hence, we keep $L_h>0$ small, such that an average of less
than $100$ particles resides in each element. On the other hand, $L_h$
must be chosen large enough to avoid motion of particles accross
multiple elements in one time step, i.e. $k_{pd} V_p \le L_h$, where
$k_{pd}>0$ is the time step size of the particle model and $V_p$ the
maximum velocity of the particles. Further, the lattice mesh is basis
for parallelization since we can guarantee that no interaction between
lattice elements which are separated by a complete layer can take
place.

\subsubsection{Configuration of the test case}\label{setup}

\begin{figure}[h]
  \includegraphics[width=\textwidth]{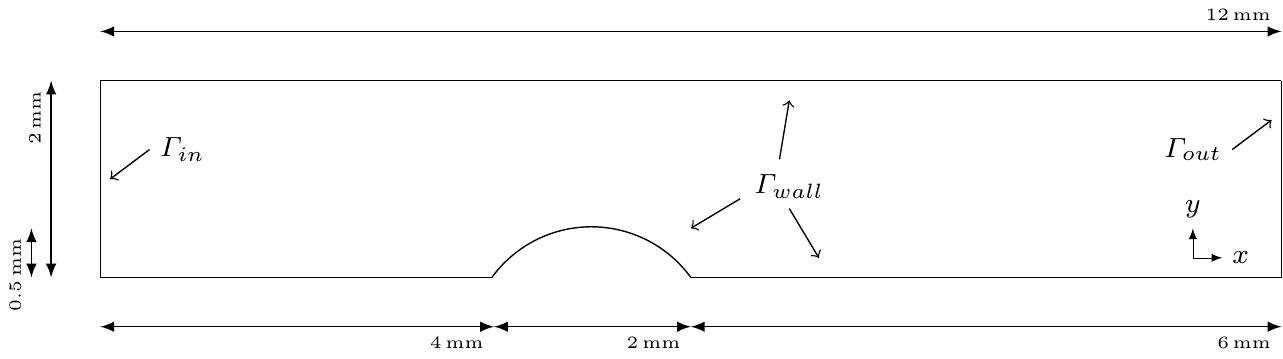}
  \caption{Geometry of the numerical examples.}
  \label{fig:domain2}
\end{figure}

We run simulations for the 2D flow in a channel with a local narrowing
of 25\% which should mimic a stenosed region of a blood
channel. Figure~\ref{fig:domain2} displays schematic geometry of the
flow domain. The size of the domain, $\unit[2]{mm}\times
\unit[12]{mm}$, is similar to the dimension of small arteries. The flow
is driven by a Dirichlet profile on the inflow boundary $\Gamma_{in}$ 
given by  
\[
\vt_{in}(x,y,t)=\frac{2-y}{2}\cdot v_{in}
\]
with $v_{in}=\unit[5]{mm/s}.$
On the wall boundary $\Gamma_{wall}$ we prescribe no-slip boundary  conditions 
$\vt=0$  and  on  the  outflow  boundary $\Gamma_{out}$  we use  the  do-nothing 
outflow condition $\mu \partial_n\vt-p\nt=0$, see~\cite{HeywoodRannacher1990}. 

The fluid viscosity is set to $\mu=\unit[3]{mg/mm\cdot s}$.
Particle and fluid densities are equal $\rho = \rho_p = 1.06\unit{mg/mm^3}$.
Due to the fact that platelets constitute less than $1\%$ 
of the blood volume~\cite{Fogelson2015} 
and the size of the domain we perform simulations 
with 165\,000  particles.

In all numerical examples the temporal step size 
for solving Navier-Stokes equations is $k_{ns}=\unit[0.005]{s}$, 
while time sub-step for particle advection is
$k_{pd}=\unit[0.00025]{s}$ such that 20 subcycles are computed in each
Navier-Stokes step.
We update the force coefficients by evaluating the neural network 
every $10$th step (i.e. twice in each Navier-Stokes step). The spatial
finite element discretization is based on quadratic elements, with a
total of 12819 degrees of freedom. 

\begin{figure}[t]
  \includegraphics[width=\textwidth]{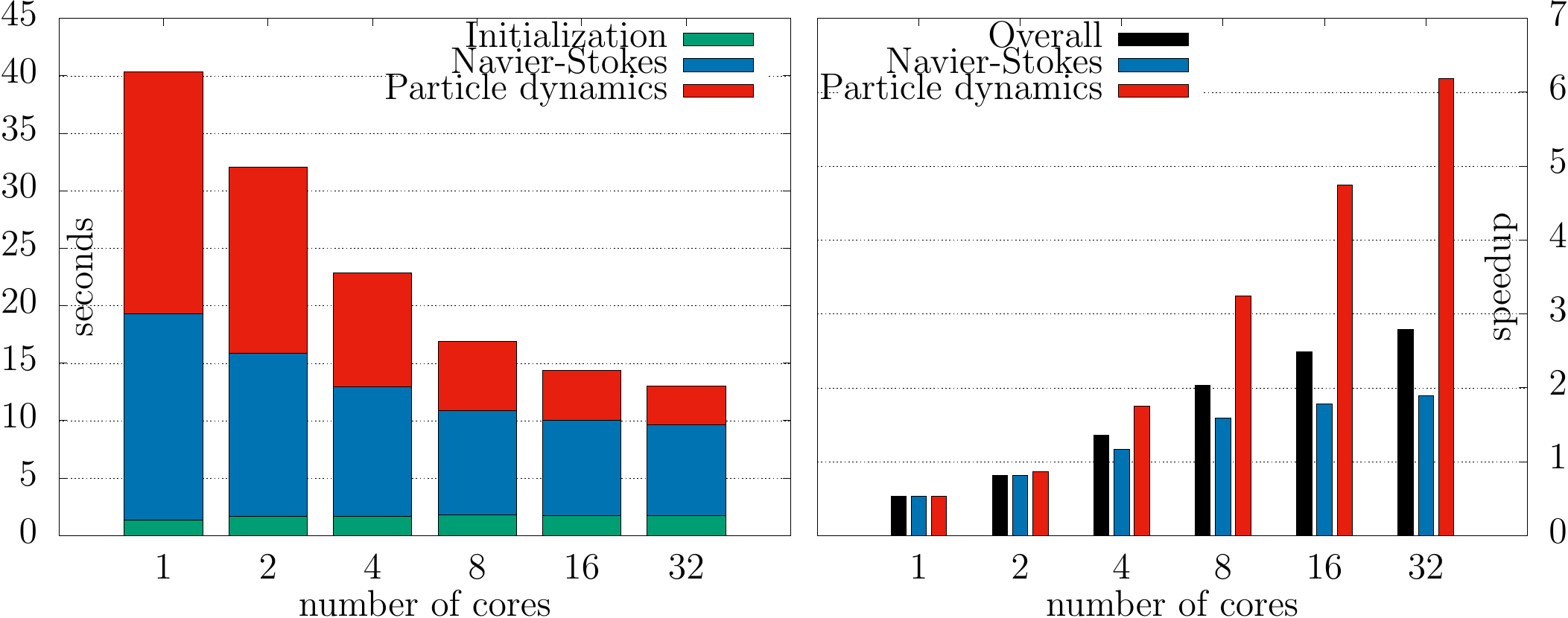}
  \caption{Left: runtime (in seconds) for the coupled Navier-Stokes
    particle dynamics simulation for an increasing number of
    cores. Right: parallel speedup for the complete simulation and for
  the Navier-Stokes finite element simulation and the particle
  dynamics simulation separately.}
  \label{fig:runtime}
\end{figure}

At the first time step we randomly seed about 165\,000 particles
distributed over the complete computational domain. Each particle is
generated with random properties, i.e. specifying 
$P=(L_x,L_y,L_z,\alpha_{top},\alpha_{bot})$ by means of the limits
indicated in~(\ref{bounds}) such that the full variety of dimensions
and shape is present. Details on the procedure for parametrization
of the particles are given in Section~\ref{sec:plateparam}.
Then, 10 iterations of the interaction loop shown in
Algorithm~\ref{algo:loop} are performed. Hence, 10 time
steps of the Navier-Stokes problem and 200
particle dynamics substeps are performed. Fig.~\ref{fig:runtime} shows
the runtime for all 200 iterations. Furthermore we
indicate the parallel speedup. These results show that the allocation
of computational time to the Navier-Stokes finite element solver and
the particle dynamics system is rather balanced. While it is 
non-trivial to get a reasonable parallel speedup for highly efficient
multigrid based finite element simulations (at least for simple 2d
problems like this Navier-Stokes testcase), the scaling of the
particle dynamics system is superior. These results demonstrate that
the number of particles is not the limiting factor for such coupled
simulations.

\begin{figure}[t]
  \includegraphics[width=\textwidth]{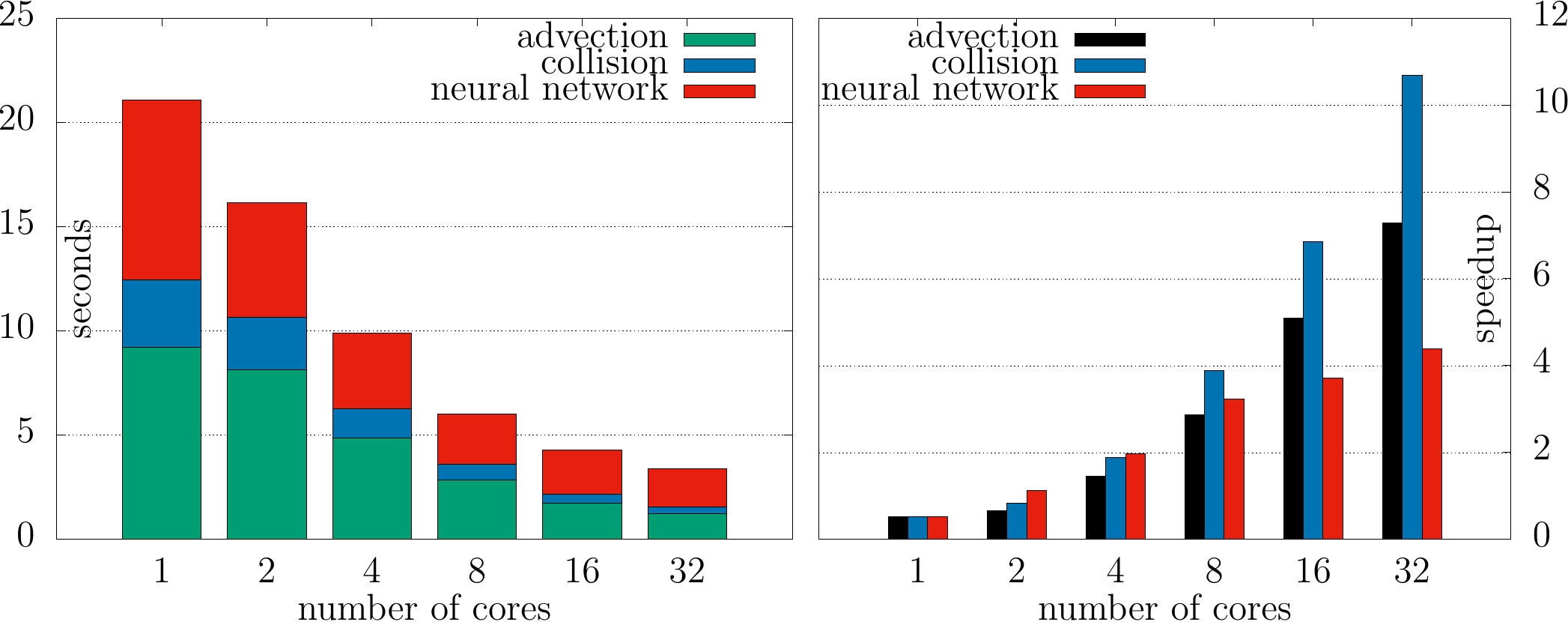}
  \caption{Left: runtime (in seconds) for the particle dynamics
    simulation (all 200 substeps). 
    Right: parallel speedup for the particle advection, handling of
    particle collisions and the neural network access for predicting
    the hemodynamical coefficients.}
  \label{fig:particles}
\end{figure}

The key feature of our coupled Navier-Stokes particle dynamics scheme
is the prediction of the hemodynamical coefficients by means of the
previously trained neural network instead of using analytical models,
which are not available, or running resolved simulations, which is not
feasible for such a large number of particles. In
Fig.~\ref{fig:particles} we give details on the computational time
spend in the different parts of the particle dynamics
system. Besides advection of the particles, the evaluation of the
neural network is dominant, alghouth we update the coefficients in
every 10th step only. Here, a more systematic study of the impact of
the update frequency should be performed. A further acceleration of
the neural network evaluation is possible by using the GPU
implementation of PyTorch. 

The transition to more realistic 3d problems will substantially
increase the effort in both parts, finite elements and particle
dynamics. For the Navier-Stokes simulation it has been demonstrated
that realistic 3d blood flow situations can be handled in reasonable
time,
see~\cite{FailerMinakowskiRichter2020,FailerRichter2020opt,FailerRichter2020}. If
the number of particles is to be substantially increased, the neural
network coupling for estimating hemodynamic coefficients should be
realized in a high performance package such as LAMMPS~\cite{LAMMPS}
that allows for an efficient GPU implementation.

\newpage
\section{Conclusions}\label{sec:conc}

Suspensions of arbitrarily-shaped particles in a fluid
are of great importance both in engineering and medical applications. 
However, the interaction of the non-spherical particles with a fluid flow 
is a complex phenomenon, even for regularly-shaped particles
in the simple fluid flows. The main difficulty lies in determining 
hydrodynamic forces experienced by a particle
due to their strong dependence on both particle shape 
and its orientation with respect to the fluid flow.

In this paper, a model is successfully derived to simulate the
motion of non-spherical particles in a non-uniform flow field,
including translation and rotation aspects. The model is  
designed to reflect platelets in a blood flow,
both in terms of particle parameters and fluid configuration.
The very good agreement of these results  obtained by the coupled
finite element / neural network / particle dynamics simulation with
state of the art documentations in literature indicates an effectiveness 
of the presented approach and hence an encouraging potential 
toward medical applications. Furthermore, the big improvement over
usual analytical interaction models is clearly seen as the neural
network based model holds for a broad range of different 
shapes at any orientation. Moreover, using neural network to identify 
the transmission of forces from fluid to the particles  
provides a possibility to adopt the model to any desired shape of particle,
making this method very promising.

We have further documented details on the scaling of the approach to
many particles, which, in 2d blood flow
simplifications, matches the typical particle density found for
thrombocytes in blood flows. The computational effort is well
balanced into the Navier-Stokes finite element part, the particle
advection and the evaluation of the neural network.

\newpage
\begin{acknowledgements}
The authors acknowledge the financial support by the Deutsche Forschungsgemeinschaft (DFG, German Research Foundation) - 314838170, GRK 2297 MathCoRe. TR acknowledges the support by the Federal Ministry of Education and Research of Germany, grant number 05M16NMA. 
\end{acknowledgements}

\section*{Conflict of interest}

The authors declare that they have no conflict of interest.

\end{document}